\documentclass[aps,twocolumn,showpacs,preprintnumbers,nofootinbib,prd,superscriptaddress,floatfix,10pt]{revtex4-2}

\usepackage{graphicx,amssymb,amsmath,amsthm,amsfonts}
\usepackage{dcolumn}
\usepackage{bm}
\usepackage{subcaption}
\usepackage{float}
\DeclareMathOperator*{\MM}{\Sigma}

\begin{document}


\title{Scalarization and superradiant instability of black hole induced by dark matter halo in  the scalar-tensor theory of gravity.}


\author{Junya Tanaka}
\email[tanaka-junya470@g.ecc.u-tokyo.ac.jp]{Your e-mail address}
\affiliation{Department of Earth Science and Astronomy, Graduate School of Arts and Science, The University of Tokyo, \\Komaba 3-8-1, Meguro-ku,Tokyo 153-8902, Japan}


\date{April 2025}
\begin{abstract}
We investigate whether a black hole(BH) surrounded by a dark matter (DM) halo has scalar hair/superradiant instability in the scalar tensor theory of gravity. In the scalar tensor theory, the coupling of matter and the scalar field creates effective mass, this effective mass causes the hairless BH  to have scalar hair  (spontaneous Scalarization). In the case of rotating BHs, it is also known that if the sign of the effective mass is positive, superradiant instability can occur instead of scalarization. Our study applies this effect to BHs with dark matter haloes. As a result, we confirmed that scalarization and superradiant instability occur in some of the parameter regions.
In the case of small haloes, the size and mass of the halo affect the strength of scalarization, but not for astronomically large haloes, where the dependence on the coupling constant $\alpha$ is stronger. In the case of superradiance, we also confirm that for small haloes, the size and mass of the halo affect the strength of the instability and the number of unstable modes.
\end{abstract}

\maketitle
\section{\label{intro}Introduction}
Scalar tensor theories of gravity are widely studied as extension of general relativity (GR). These theories involve one or more scalar fields that are non-minimally coupled to metric. When there is one scalar field, the action is generally described by the Jordan frame (J frame) \cite{Fujii:2003pa,faraoni2004cosmology}:
\begin{align}\label{Jordanframe}
S=\frac{1}{16\pi G}\int d^4x \sqrt{-g}(F(\phi)R-Z(\phi)g^{\mu\nu}\partial_{\mu}\phi \partial_{\nu}\phi \nonumber \\
-U(\phi))+S(\Psi_m;g_{\mu \nu}))
\end{align}
Here $R$ is Ricci Scalar, $g_{\mu\nu}$ is the spacetime metric, $\phi$ is the scalar field, and $G$ is a constant. Theories with non-minimal couplings arise from the higher dimensional theory such as string theory and this feature is always included in scalar tensor theories. Lagrangian (\ref {Jordanframe}) is a general representation of scalar tensor theory, but by determining $F,Z,U$, we can determine the specific theory class. For example, for Brans-Dicke, $F=\phi$, $Z=\omega_/\phi$, $U=0$ \cite{PhysRev.124.925}.\par
These theories must be consistent with GR in the weak field limit. Tests in the solar system and at local scales have verified GR with high accuracy. Therefore, a mechanism to screen scalar fields at these scales is necessary, and various screening mechanisms have been proposed \cite{2004PhRvL..93q1104K,2010PhRvL.104w1301H}. Therefore, in order to verify a scalar tensor theory, it is necessary to discuss the phenomena in a strong gravitational environment such as around BH\par

Generally, in modified gravity theories with scalar fields, the spacetime is  different from that of GR, and scalar fields can have notrivial configurations around BHs. However, there is a subclass among them that has the same spacetime as that of GR when the scalar field is constant. One might think that such a spacetime would be indistinguishable from GR's, but the difference in perturbation dynamics could be probed  \cite{2018PhLB..781..728M}. In fact, the scalar tensor theory are reported to deviate from GR \cite{2008PhRvL.101i9001B,2011PhRvL.107x1101C,2012PhRvD..85j2003Y}.\par
In scalar-tensor theories, it is known that around BHs, the scalar field is a constant in the case of stationary, asymptotically flat spacetime (isolated BHs), and that the spacetime around BHs is described by the Kerr-Newmann family \cite{2012PhRvL.108h1103S}. However, contrary to the well-known no hair theorem \cite{misner2017gravitation}, there are usually accretion disks, gases and other matter around astronomical BHs. Thus, if a matter field exists around the BH, the scalar field generally takes on a non-trivial configuration.\par
According to \cite{2013PhRvD..88d4056C}, there is a subclass of theories that have a configulation in which the scalar field is constant around the GR solution BH even in the presence of matter. At the same time, however, even in such a subclass, the scalar field takes on a nontrivial configuration due to perturbation effects, forcing the black hole to have scalar hair. This effect is called "spontaneous scalarization" and (at the linear level) is a manifestation of tachyonic instability \cite{2024RvMP...96a5004D}. This instability is caused by the square of the negative effective mass of the scalar field, which is created by the coupling of the scalar field to the trace of the energy-momentum tensor of the matter field. This scalarization mechanism is one of the models belonging to the DEF (Damour and Esposito-Fa$\Grave{r}$ese) model \cite{PhysRevLett.70.2220}. The scalarization of neutron stars by the DEF model has been studied in detail \cite{1997PThPh..98..359H,1998PhRvD..58f4019N}.\par
If the sign of the squared term of the effective mass is positive,  spontaneous superradinace effect can occur around a rotating BH instead of scalarization. The superradiant phenomenon is the amplification of waves incident on a rotating BH if the superradiant condition \cite{PismaZhETF.14.270,Zeldovich1972AmplificationOC}
\begin{equation}\label{superradint condition}
\omega<m \Omega
\end{equation}
is satisfied. Here, $\omega$ is the wave frequency, $m$ is the azimuthal number, and $\Omega$ is the angular velocity of the BH. Superradiant instability can occur when there is a mechanism confining this amplified wave outside the BH. For example, when the spacetime is asymptotically AdS spacetime \cite{2004PhRvD..70h4011C,2006PhRvD..74d4008C,PhysRevD.80.084020} the AdS boundary confine  fluctuations, or when the scalar field has its own mass \cite{2017PhRvL.119d1101E,2018PhRvL.121m1104E}, it play the role of a mirror. In the scalar-tensor theory, superradiant instability can occur when the effective mass $\mu^2_{eff}$ generated by the coupling of scalar field and matter replaces the mass of the scalar field.  Superradiant instability by this mechanism has been studied in the case of Kerr-de Sitter BH\cite{2014JHEP...08..011Z} and Kerr BH with accretion disk \cite{2022PhRvD.106b4007L}.\par
In current cosmology, the existence of dark matter (DM) is suggested by various data, such as galaxy rotation curves (RCs) and the large-scale structure of the universe, and it is known that the existence of DM is important in shaping the structure of the universe. The nature of DM is still unknown, but the most widely accepted theory is that of cold dark matter (CDM) \cite{1996ApJ...462..563N,1997ApJ...490..493N}. However, CDM has some issues, for example, the problem of the density profile of the galactic center \cite{1995ApJ...447L..25B}. As a result, various alternative theories of DM, such as scalar field dark matter (SFDM) \cite{Matos2004,2011JCAP...05..022H,2018JPhCS1010a2005F}, have been proposed to solve these problems. At the galactic level, galaxy rotation curves indicate that a dark matter halo extends down to the central BH of a galaxy. It is also known that there could be a mini-halo of dark matter around primordial black holes in the early universe \cite{2023MNRAS.520.4370D,2023PhRvD.107h3013Z}. Thus, dark matter is thought to exist around BHs in the astrophysical environment. In other words, it is possible that the dark matter around BHs couples to the scalar field as the matter, causing spontaneous scalarization or superradiance instability of the BH. \par In this study, we analyze in detail the spontaneous scalarization and superradiant instability induced by dark matter around  BH. If scalarization/superradiant instability occurs around the BH, it should be a existential proof of the scalar field and the scalar tensor theory. However, the identity of DM (as a particle) is not yet known, and the model of the dark matter halo has not yet been identified. Therefore, in this study, we use two representaive profiles, i.e, NFW model and SFDM model. The NFW model represents a halo of cold dark matter. The SFDM model is one in which the dark matter is a scalar field and the halo is a Bose-Einstein condensate. The analytical forms of the metrics representing their halos and BHs are already known for these two models \cite{2018JCAP...09..038X}.
This paper is organized as follows. In section \ref{DMsec}, we review the DM halo metric around BH. In section \ref{STTsec}, we look at how effective mass arises in scalar tensor theory, and then in section \ref{Scalsec} we investigate whether scalarization occurs in the specific halo. Here we see the difference in the dependence of the halo parameters in the small and large halos. In section \ref{SIsec}, we study how the effective mass created by the halo triggers superradiant instability. We also see how the halo parameter affects the instability. The section \ref{Consec} is devoted to conclusion and discussion, and the Appendices provides details on the calculation of  the effective mass and the separation of the Klein-Goldon equation.\par
In this paper, we use the units that $G=c=1$ and the $(-, +, +, +)$ signaure for the metric.
\section{\label{DMsec}Dark matter halo models}
\subsection{Metrics}
Here we will look at the spacetime of the dark matter halo containing a BH. We need to know the metric of that spacetime for the calculation of scalarization; the metric representing the spacetime of the dark matter halo containing a BH is already known in analytic form for several halo models \cite{2018JCAP...09..038X}. For CDM model, the line element is 
\begin{align}
ds^2=&-\left[\Big[1+\frac{r}{R_s}\Big]^{-{8\pi \rho_c R_s^3}/{ r}}-\frac{2M}{r}\right]dt^2 \nonumber \\
&+\left[\Big[1+\frac{r}{R_s}\Big]^{-{8\pi \rho_c R_s^3}/{ r}}-\frac{2M}{r}\right]^{-1}dr^2\nonumber \\
&+r^2(d\theta^2+\sin^2\theta d\phi^2) .
\end{align}
Here, $\rho_c$ is the density when the halo collapses, $R_s$ is the characteristic radius.\par
The derivative of the metric component are
\begin{align}
f(r)&\equiv\Big[1+\frac{r}{R_s}\Big]^{-{8\pi  \rho_c R_s^3}/{r}}-\frac{2M}{r}, \label{cdmmetric}\\
f'(r)&=\Big[1+\frac{r}{R_s}\Big]^{-{8\pi \rho_c R_s^3}/{r}} \Big[ \frac{8\pi  \rho_c R_s^3}{r} \log\Big[1+\frac{r}{R_s}\Big]\nonumber \\
& -\frac{8\pi \rho_c R_s^3}{ r(r+R_s)}  \Big]+\frac{2M}{r^2} ,\\
f''(r)&=\Big[1+\frac{r}{R_s}\Big]^{-{8\pi \rho_c R_s^3}/{r}}\Big[ -\frac{16\pi \rho_c R_s^3}{r^3}\log\Big[1+\frac{r}{R_s}\Big] \nonumber \\
&+ \frac{8\pi \rho_c R_s^3}{r^2(r+R_s)}-\frac{8\pi \rho_c R_s^3}{r}\frac{1}{R_s} \Big[\frac{1}{(r+R_s)^2}-\frac{1}{r^2}\Big] \nonumber \\
&+\left(\Big[1+\frac{r}{R_s}\Big]^{-{8\pi \rho_c R_s^3}/{r}}\right)'\Big[ \frac{8\pi \rho_c R_s^3}{r} \log\Big[1+\frac{r}{R_s}\Big] \nonumber \\
&-\frac{8\pi \rho_c R_s^3}{r(r+R_s)} \Big]-\frac{4M}{r^3} ,
\end{align}
when commas mean derivative with respect to $r$.
For SFDM model,
\begin{align}
ds^2=&-\left[\exp\Big(\frac{8\rho_c R^2}{ \pi} \frac{\sin(\pi r /R)}{\pi r/R}\Big) -\frac{2M}{r}\right]dt^2 \nonumber \\
&+\left[\exp\Big(\frac{8\rho_c R^2}{\pi} \frac{\sin(\pi r /R)}{\pi r/R}\Big) -\frac{2M}{r}\right]^{-1}dr^2\nonumber \\
&+r^2(d\theta^2+\sin^2\theta \phi^2) .
\end{align}
Here, $\rho_c$ is the central density, $R=\pi/k$ is the radius that the presure and the density vanish, and $k$ is the Compton wave number of DM particle.
Derivatives of the metric are

\begin{align}
f(r)=&\exp\Big(-\frac{8\rho_c R^2}{\pi} \frac{\sin(\pi r /R)}{\pi r/R}\Big) -\frac{2M}{r} ,\label{sfdmmetric}\\
f'(r)=&-\frac{8\rho_c R^2}{\pi}\frac{\pi}{R} \left[\frac{\cos(\pi r/R)}{\pi r/R}-\frac{\sin(\pi r/R)}{(\pi r/R)^2} \right]\nonumber \\
&\exp\Big(-\frac{8\rho_c R^2}{\pi}\frac{\sin(\pi r /R)}{\pi r/R} \Big)+\frac{2M}{r^2},\\
f''(r)=&\Big(-\frac{8\rho_c R^2}{\pi}\Big)^2 \left[\frac{\cos(\pi r/R)}{\pi r/R}-\frac{\sin(\pi r/R)}{(\pi r/R)^2} \right]^2\nonumber \\
& \exp\Big(-\frac{8\rho_c R^2}{\pi}\frac{\sin(\pi r /R)}{\pi r/R}\Big)\nonumber \\
&-\frac{8\rho_c R^2}{\pi}\Big(\frac{\pi}{R} \Big)^2 \Big[ -\frac{\sin(\pi r/R)}{\pi r/R}-2\frac{\cos(\pi r/R)}{(\pi r/R)^2} \nonumber \\
&-2\frac{\sin(\pi r/R)}{(\pi r/R)^3} \Big]\exp\Big(-\frac{8\rho_c R^2}{\pi}\frac{\sin(\pi r /R)}{\pi r/R}\Big)-\frac{4M}{r^3} .
\end{align}

\begin{figure*}[th]
\centering
\begin{minipage}{1.0\columnwidth}
\centering
\includegraphics[width=\columnwidth]{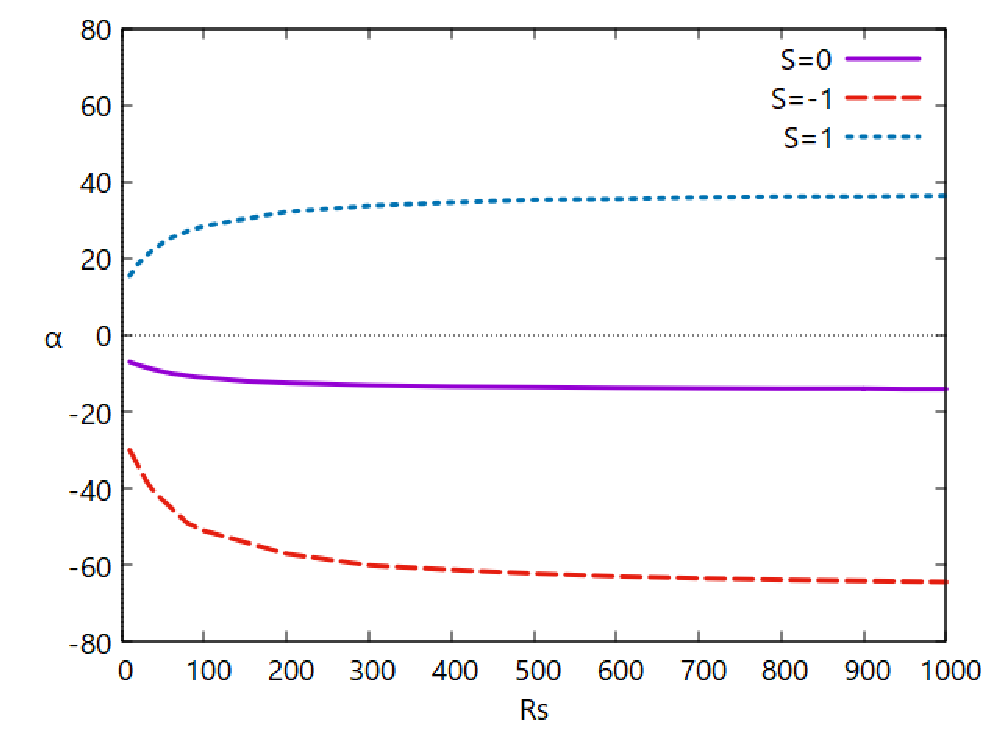}
\caption{\label{CDMSplane}The contour curves of $S$ in $R_s-\alpha$ plane. Each line in the figure represents $S=-1.0, 0, 1.0$, respectively (CDM, $1\le R_s \le1000$, $M_{_\mathrm{DM}}= 1.0, \ell=0$).}
\end{minipage}
\begin{minipage}{1.0\columnwidth}
\centering
\includegraphics[width=\columnwidth]{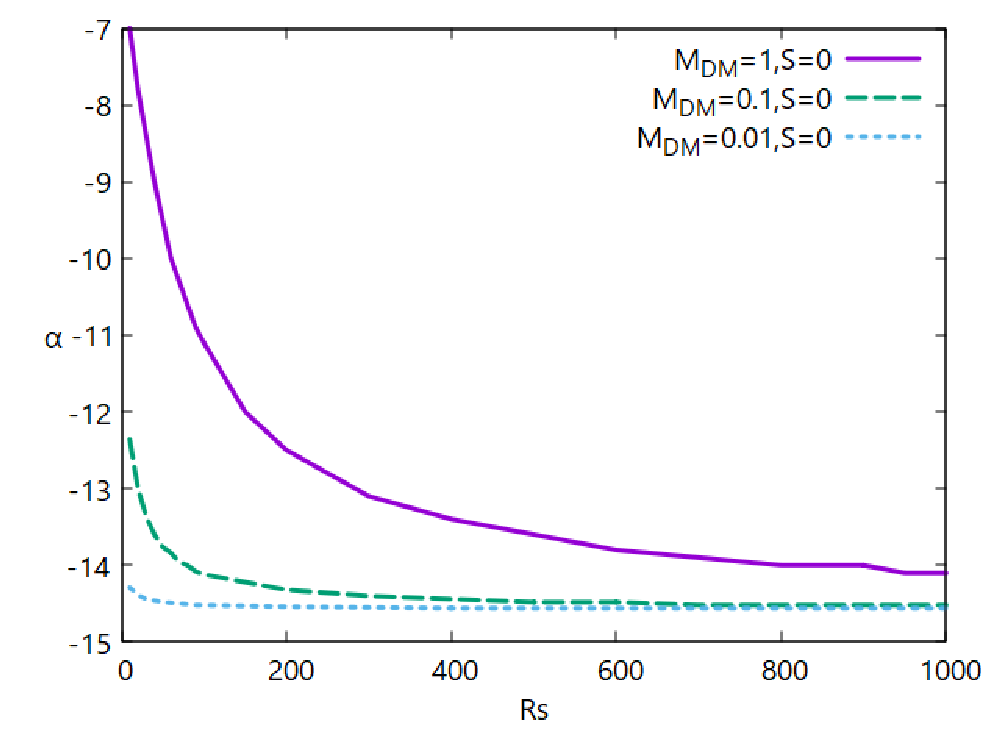}
\caption{\label{CDMSplaneMDM}The lines of $S=0$ for $M_{_\mathrm{DM}}= 0.01, 0.1$ and $1.0$ on the $R_s-\alpha$ plane (CDM,  $1\le R_s \le1000$, $\ell=0$).}
\end{minipage}
\end{figure*}

\subsection{Energy momentum tenosr}
For a given metric form of spacetime, energy-momentum tensor of the dark matter halo is defined by Einsten equation. The metric for a general spherically symmetric spacetime is
\begin{equation}
ds^2=-f(r)dt^2+\frac{1}{g(r)}dr^2+r^2(d\theta^2+\sin^2\theta \phi^2) .
\end{equation}
By computing the Einstein equations 
\begin{equation}
R^{\nu}_{\mu}-\frac{1}{2} \delta^{\nu}_{\mu}R=\kappa T^{\nu}_{\mu} ,
\end{equation}
from the metric, each component of the energy-momentum tensor can be derived:
\begin{align}
\kappa T^t_t=&g(r)\left(\frac{1}{r}\frac{g'(r)}{g(r)}+\frac{1}{r^2}\right)-\frac{1}{r^2} ,\\
\kappa T^r_r=&g(r)\left(\frac{1}{r}\frac{f'(r)}{f(r)}+\frac{1}{r^2}\right)-\frac{1}{r^2} ,\\
\kappa T^{\theta}_{\theta}=&\kappa T^{\phi}_{\phi}=\frac{1}{2} g(r)\Big[\frac{f''(r)f(r)-f'^2(r)}{f^2(r)}+\frac{1}{2}\frac{f'^2(r)}{f^2(r)}\nonumber \\
&+\frac{1}{r}\Big(\frac{f'(r)}{f(r)}+\frac{g'(r)}{g(r)}\Big)+\frac{f'(r)g'(r)}{2f(r)g(r)}\Big].
\end{align}
In the case of $f(r)=g(r)$,
\begin{equation}
\kappa T=2f(r)\left(\frac{1}{r}\frac{f'(r)}{f(r)}+\frac{1}{r^2}\right)-\frac{2}{r^2}+\frac{1}{2}f''(r)+\frac{1}{r}f'(r) .
\end{equation}


\section{\label{STTsec}Scalar-Tensor theory}
In this section, by following \cite{2013PhRvD..88d4056C}, we see the mechanism of the spontaneous scalarization/superradiant instability induced by the scalar field-matter coupling .We start from Jordan frame (\ref{Jordanframe}).
By performing a conformal transformation and field redifinition, we have
\begin{align}
g_{\mu \nu}&=F(\phi) g_{\mu \nu}^E ,\\
\Phi(\phi)&=\frac{1}{\sqrt{4 \pi}} \int d\phi \left[ \frac{F'(\phi)^2}{F(\phi)^2} + \frac{1}{2} \frac{Z(\phi)}{F(\phi)}\right]^{1/2} ,\\
A(\Phi)&=F^{-1/2}(\phi) ,\\
V(\Phi)&=\frac{U(\phi)}{F^2(\phi)} .
\end{align}
We can remove non-minimal couplings and move to an another conformal frame (Einstein frame):
\begin{align}
S=\frac{1}{16\pi G}\int d^4x \sqrt{-g^E}\Big(\frac{R^E}{16 \pi}-\frac{1}{2}g^E_{\mu\nu}\partial^{\mu}\Phi \partial^{\nu}\Phi-V(\Phi)\Big) \notag \\
+S(\Psi_m;A(\Phi)g_{\mu \nu}^E) .
\end{align}
The field equations in the Einstein frame, obtained by variating $\Phi$ and $g_{\mu\nu}^E$, are
\begin{align}
G_{\mu \nu}&=8\pi \left(T^E_{\mu  \nu}+\partial_{\mu}\Phi \partial_{\nu}\Phi -\frac{g^E_{\mu \nu}}{2}(\partial \Phi)^2   \right)-\frac{g^E_{\mu \nu}}{2}V(\Phi) \\
\square^E \Phi&=-\frac{A'(\Phi)}{A(\Phi)}T^E+\frac{V'(\Phi)}{16\pi} .
\end{align}
Here,
\begin{eqnarray}
T^{\mu E}_{\nu}=A^4 T^{\mu}_{\nu} ,\\
T^E_{\mu \nu}=A^2 T_{\mu \nu} , \\
T^E=A^4T ,
\end{eqnarray}
are the energy-momentum tensor of matter field and its trace .
Originally, the spacetime should have been a GR solution with a constant scalar field $\Phi_0$, 
so we assume that the scalar field is expanded near $\Phi \sim \Phi_0$,
\begin{align}
V(\Phi)=\sum_{n=0}V_n(\Phi-\Phi_0)^n ,\\
A(\Phi)=\sum_{n=0}A_n(\Phi-\Phi_0)^n .
\end{align}
At the first order of expansion of $\varphi=\Phi-\Phi_0 \ll 1$, the fields equations are \cite{2012PhRvD..85j2003Y}
\begin{align}
&G_{\mu \nu}=8\pi \left(T^E_{\mu  \nu}+\partial_{\mu}\Phi_0 \partial_{\nu}\Phi_0 -\frac{g^E_{\mu \nu}}{2}(\partial \Phi_0)^2   \right)-\frac{g^E_{\mu \nu}}{2}V_0\notag\\
&+8\pi \left(\partial_{\mu}\Phi_0 \partial_{\nu}\varphi_0 +\partial_{\mu}\varphi_0 \partial_{\nu}\Phi_0-g^E_{\mu \nu}\partial_{\mu} \Phi_0 \partial^\mu \varphi   \right)-\frac{g^E_{\mu \nu}}{2}V_1 \varphi, \\
&\square^E \Phi_0+\square^E \varphi =-\frac{A_1}{A_0}T^E+\frac{V_1}{16\pi}+\frac{V_2 \varphi}{8\pi}\notag\\
&\quad \quad\quad\quad\quad\quad+\varphi T^E \left(\frac{A_1^2}{A_0^2}-2\frac{A_2}{A_0}\right).
\end{align}
$V_0$ term is related to the cosmological constant $\Lambda \equiv V_0/2$. We set $V_0=0$ because of the assumption of asymptotic flatness. Also, we set the first term $V_1=0$ of the potential. The $V_2$ is related to the standard mass term. When $A_1 \neq0$, the BH solution surrounded by matter will always have a nontrivial scalar field profile. Unlike the GR one, this configuration depends on $A_1$ and the matter field.\par

No solution exists if the matter field exists and $\Phi_0= const.$ except when $A_1=0$. In the case of $A_1=0$, the field equations have GR solutions with stationary background scalar fields. Nevertheless, nontrivial scalar fields arise in hairless BH due to perturbative effects.  \par 
If $A_1=0$ and $V_0=0=V_1$, in the first order of $\varphi$, then the KG equation should be

\begin{equation}\label{KG}
\left[\square^E -\frac{V_2 \varphi}{8\pi}+2\frac{A_2}{A_0}T^E \right]\varphi \equiv [\square^E -\mu_s^2]\varphi =0 ,
\end{equation}
where
\begin{equation}\label{effective mass}
\mu_s^2\equiv \frac{V_2 \varphi}{8\pi}-2\frac{A_2}{A_0}T^E ,
\end{equation}
and
\begin{equation}\label{coupling}
\alpha \equiv  \frac{A_2}{A_0} ,
\end{equation}
where $\mu_s^2$ is the effective mass of the scalar field resulting from the coupling of the scalar field to the matter field. $V_2$ is a term related to the standard mass, which we now ignore for the sake of simplicity, since it does not contribute significantly to the discussion in this paper ($V_2=0$). Depending on the signs of $A_2$ and $T^E$, $\mu_s^2$ can take both positive and negative values. When $\mu_s^2$ is positive, superradiance instability may appear if BH is rotating, and when $\mu_s^2$ is negative, spontaneous scalarization appears as tachyonic instability.


\section{Scalarization}
\label{Scalsec}
Since the dark matter halo spacetime (CDM and SFDM) including non-rotating BH are spherically symmetric, we can decompose the scalar field as $\varphi(t,r,\theta,\phi)=\Sigma_{\ell m}\frac{1}{r}\phi_{\ell m}e^{-i\omega t}Y_{\ell m}$, then Eq.(\ref{KG}) can be written as
\begin{align}
&\frac{d^2 \Psi(r)}{dr^2_*}+[\omega^2-V(r)]\Psi(r)=0 \label{schrodinger} ,\notag\\
&V(r)=f(r) \times \left(\frac{\ell(\ell+1)}{r^2}+\frac{f'(r)}{r}+\mu_s^2 \right) .
\end{align}
where $r_*$ is the tortoise coordinate (defined as $dr/dr_*=1/f$), $f$ is (tt)-componet of the metric which is Eq. (\ref{cdmmetric}) for CDM or eq.(\ref{sfdmmetric}) for SFDM.
$\omega$ is a characteristic frequency and has real and imaginary parts as $\omega=\omega_R+i\omega_I$. Eq.(\ref{schrodinger}) is an eigenvalue equation with $\omega$ as an eigenvalue. For this eigenvalue equation to have unstable modes, it is sufficient that $\omega_I>0$.
From a well-known result in quantum mechanics \cite{1995AmJPh..63..256B}, the condition for instability is
\begin{align}
&\int_{2M}^{\infty} \frac{V}{f} \mathrm{d}r \notag \\
=&\int_{2M}^{\infty}\left(\frac{\ell(\ell+1)}{r^2}+\frac{f'(r)}{r}-2\alpha T^E   \right)\mathrm{d}r<0 . \label{instability}
\end{align}
\par
We use this instability condition (\ref{instability}) to test whether the dark matter halo causes spontaneous scalarization around the BH. Also, the value of $\alpha$ is observationally limited by binary-pulsars \cite{2024RvMP...96a5004D}. Taking this into account, we further examine whether or not the DM halo causes scalarization in a realistic manner for some specific halos. We used the halo parameters described in \cite{2012MNRAS.422..282R} and \cite{2018JPhCS1010a2005F}.
\par

\begin{figure*}[ht]
\centering
\begin{minipage}{1.0\columnwidth}
\centering
\includegraphics[width=\columnwidth]{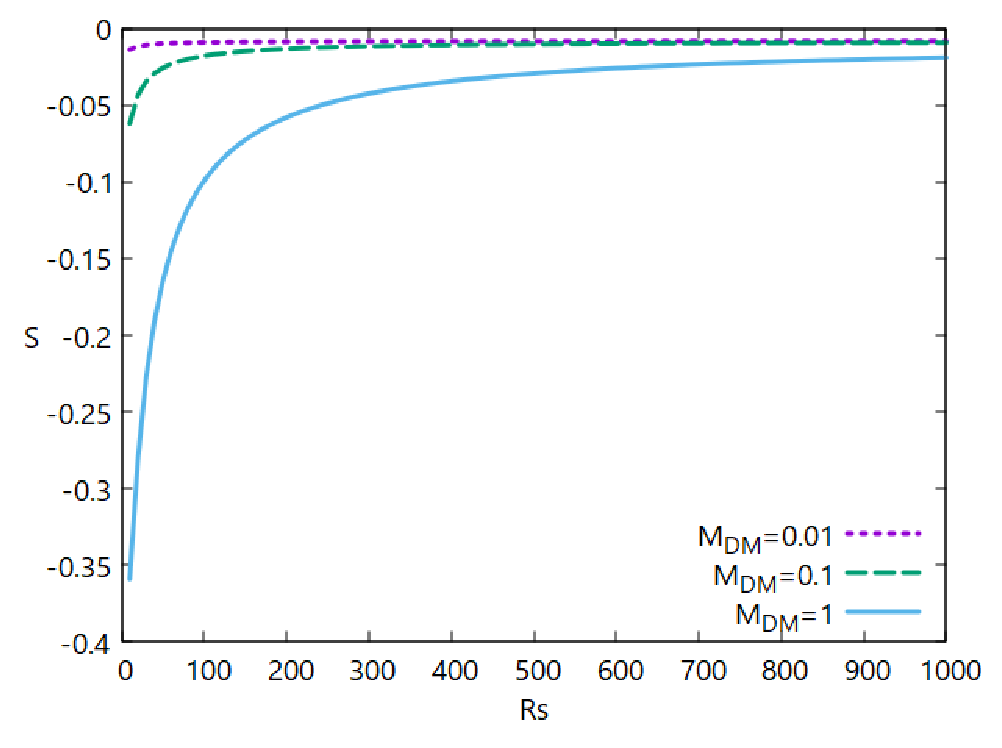}
\caption{\label{CDMalpha-15}The value of S with changing $R_s$ from $1.0$ to $1000$ for $M_{_\mathrm{DM}}= 0.01, 0.1$ and $1.0$ and fixing $\alpha=-15$ (CDM, $\ell=0$).}
\end{minipage}
\begin{minipage}{1.0\columnwidth}
\centering
\includegraphics[width=\columnwidth]{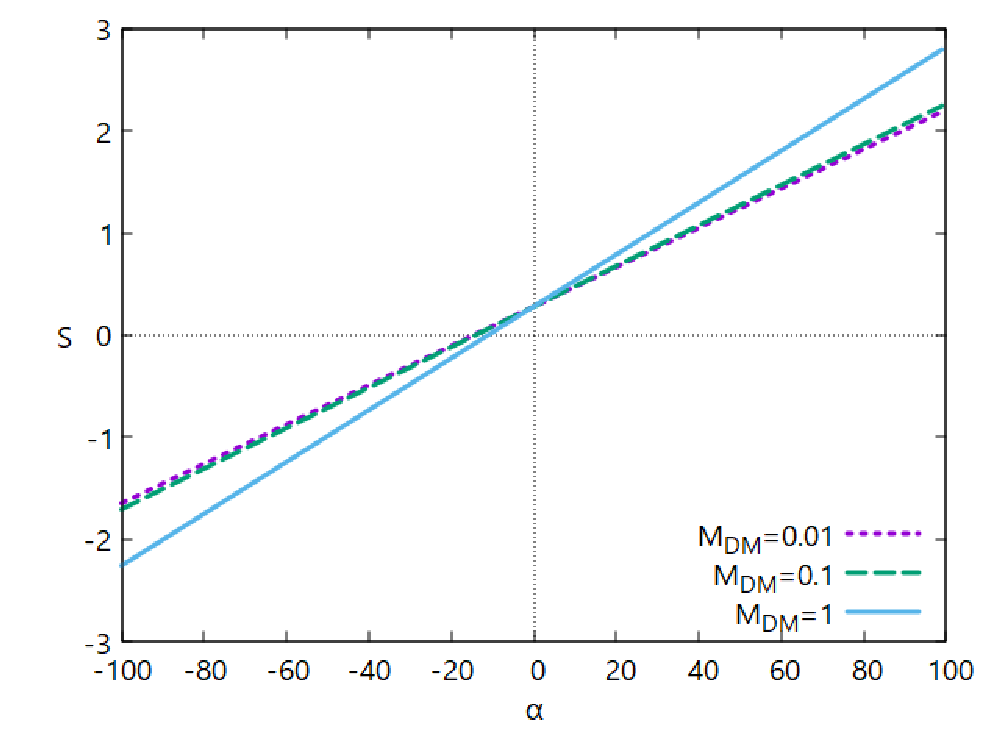}
\caption{\label{CDMMDM0.01-1R100}The value of S with changing $\alpha$ from $-100$ to $100$ for $M_{_\mathrm{DM}}= 0.01, 0.1$ and $1.0$ and fixing $R_s=100$ (CDM, $\ell=0$).}
\end{minipage}
\end{figure*}

\begin{figure*}
\centering
\begin{minipage}{1.0\columnwidth}
\centering
\includegraphics[width=\columnwidth]{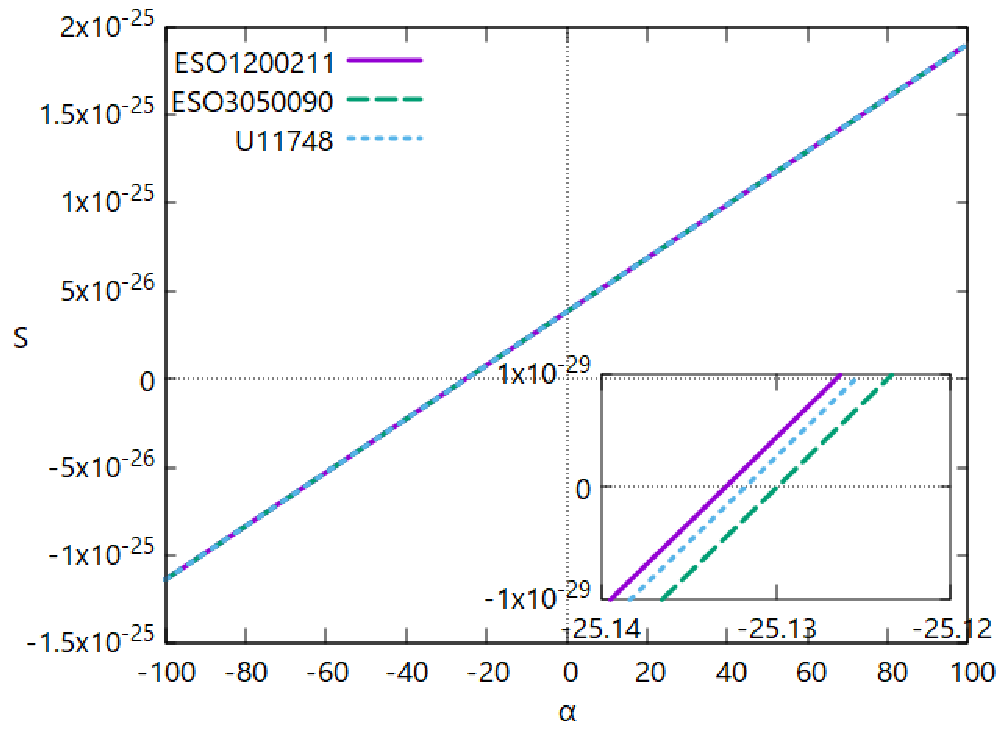}
\caption{\label{cdmcompare}The values of $S$ with changing $\alpha$ for ESO1200211, ESO3050090 and U11748 (CDM, $\ell=0$). The inset figure is an enlarged view of the area where each line is S=0.}
\end{minipage}
\begin{minipage}{1.0\columnwidth}
\includegraphics[width=\columnwidth]{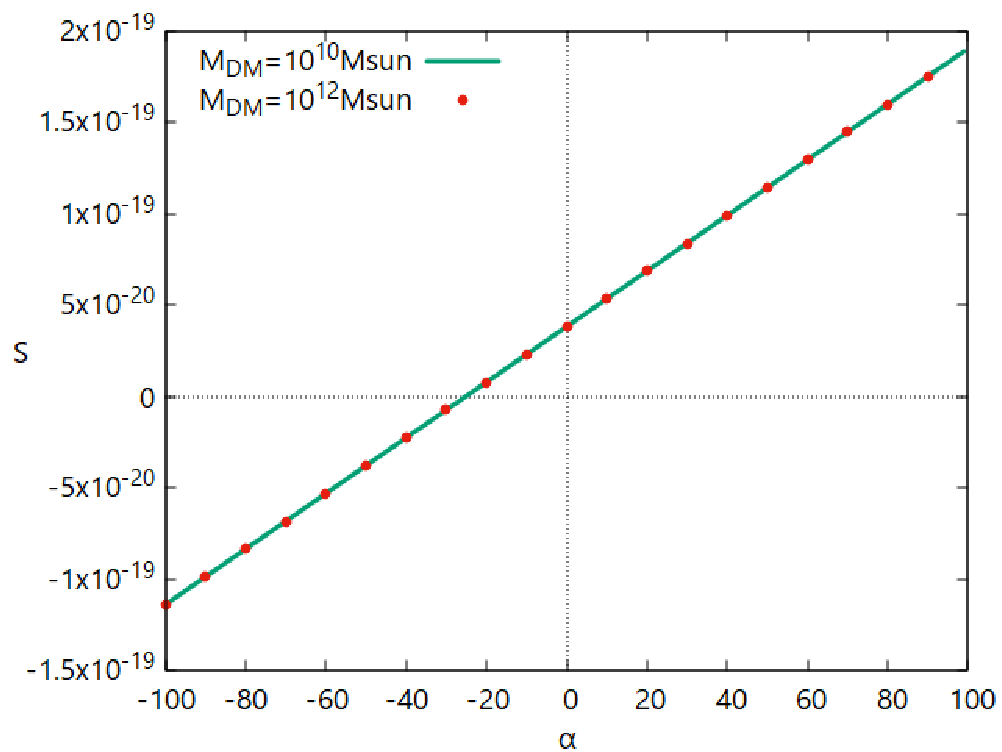}
\caption{\label{CDMDMmass}The value of $S$ with $M_{_\mathrm{DM}}=1.0 \times 10^{10}M_{\odot}$ and $1.0 \times 10^{12}M_{\odot}$,changing $\alpha$ (CDM, $\ell=0$,$R_s=10$kpc, $M=1.0 \times 10^6M_{\odot}$).}
\end{minipage}
\end{figure*}

\begin{figure}
\includegraphics[width=\columnwidth]{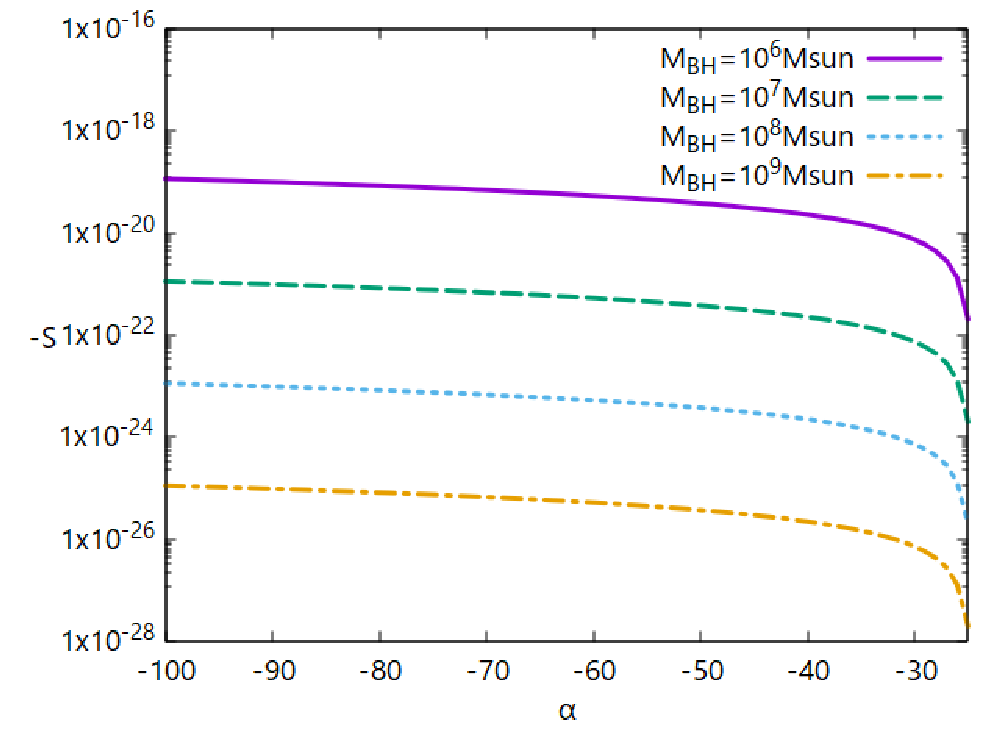}
\caption{\label{CDMBHmass}The value of $S$ with $M=1.0 \times 10^6M_{\odot} -10^{9}M_{\odot}$, changing $\alpha$ (CDM,$\ell=0$,$R_s=10$kpc, $M_{_\mathrm{DM}}=1.0 \times 10^{10}M_{\odot}$).}
\end{figure}

\subsection{CDM}
We will first look at the CDM model. At first, we will examine general properties without specifying a specific halo model. That is, we deal with smaller halos than actual astrophysical halos. We take the unit system $c=G=M=r_g(=GM/c^2)=1$ in this section, except for the later chapter on adaptation to specific haloes. From the instability condition (\ref{instability}) in the previous section ,scalarization occur in the area that
\begin{equation}\label{area}
S=\int_{2M}^{\infty}\left(\frac{\ell(\ell+1)}{r^2}+\frac{f'(r)}{r}-2\alpha T^E   \right)\mathrm{d}r ,
\end{equation}
is negative.

The condition (\ref{instability}) depends on the scalarization parameter $\alpha$ and the space-time metrics $f(r)$ and $T^E$. Since $f(r)$ and $T^E$ are determined by the halo parameters $R_s$ and $\rho_c$, scalarization depends on the halo structure. Therefore, we investigated the parameter region that satisfies the condition of scalarization by changing $\alpha$, $R_s$, and $\rho_c$. $\rho_c$ and $R_s$ are related by the NFW profile:

\begin{equation}
\rho_{_\mathrm{NFW}}=\frac{\rho_c}{(r/R_s)(1+r/R_s)^2}.
\end{equation}

Here, We fix the mass of the dark matter halo $M_{_\mathrm{DM}}$ and vary $R_s$ to determine $\rho_c$ from the relation 
\begin{align}\label{nfwprofile}
M_{_\mathrm{DM}}&=\int_{2M}^{R_s}\mathrm{d}r4\pi r^2 \rho_{NFW}\notag\\
           &=\int_{2M}^{R_s}\mathrm{d}r \frac{4 \pi r^2 \rho_c}{(r/R_s)(1+r/R_s)^2} .
\end{align}
\par
Fig.\ref{CDMSplane} shows $S=-1.0, 0, 1.0$ contour curves for $M_{_\mathrm{DM}}= 1.0$, and fig.\ref{CDMSplaneMDM} shows $S=0$ for $M_{_\mathrm{DM}}=0.01, 0.1, 1.0$ in the $R_s-\alpha$ plane. The scalarization occurs in the region below the $S=0$ line in these figures. The value of $S$ becomes negative in the $\alpha<0$ region for the scalarization. This is a same trend as ordinary materials. This is because $T^E \sim -\rho$ (where $\rho$ is the mass energy density) in ordinary matter such as perfect fluids and $f'(r)>0$ in Schwarzshild spacetime, it is often desirable to have $\alpha<0$ in order for $S<0$, (however, this is not the case for exotic materials for which $T^E>0$) .
\par 
 Fig.\ref{CDMalpha-15} show the values of $S$ with changing $R_s$ from $1$ to $1000$ for $M_{_\mathrm{DM}}= 0.01, 0.1$ and $1.0$ and fixing $\alpha=-15$. Fig.\ref{CDMMDM0.01-1R100} show the value of S with changing $\alpha$ from$-100$ to $100$ for $M_{_\mathrm{DM}}= 0.01, 0.1$ and $1.0$ and fixing $R_s=100$. The larger $M_{_\mathrm{DM}}$ is, the larger the absolute value of $S$ is. This can be rephrased as the scalarization is likely to occur for larger central densities, since the  mass of the halo $M_{_\mathrm{DM}}$ is considered to be larger for the same value of $R_s$ from Eq.(\ref{nfwprofile}) as $M_{_\mathrm{DM}}$ increases. This property, that the  scalarization is more likely to occur when the central density is larger, is also the same as for ordinary materials. This can be also seen from the fact that the value of S falls in the range where $R_s$ is small and the modulus of negative $\alpha$ is large.

\subsection{CDM for actual halo}

\begin{table}
\caption{table1}
\label{table1}
\begin{ruledtabular}
\begin{tabular}{cccccccc}
Halo &$\rho_c(\times 10^{-3} M_{\odot}pc^{-3})$&$R_s(kpc)$ \\ \hline
ESO1200211&2.45&5.7 \\
ESO3050090&0.0328&705.67 \\
U11748&204.58&5.53 \\ \hline
\end{tabular}
\end{ruledtabular}
\end{table}

 In this section, we investigate scalarization for DM haloes in the actual galaxies. As examples, we use halos of low surface brightness (LSB) galaxies because LSB galaxies are composed mostly of DM \cite{1997ASPC..117...39D}. The $\rho_c$ and $R_s$ are different for each galaxies. These parameters are obtained by fitting to the observed data. The haloes used as  examples are listed in the table \ref{table1}. We consider that central BH masses are the average value of the central BH of LSB galaxies, $M=5.62 \times 10^6  M_{\odot}$ \cite{2016MNRAS.455.3148S}.
\par

Fig.\ref{cdmcompare} shows the values of $S$ for ESO1200211, ESO3050090, and U11748 as $\alpha$ is varied. The values of $S$ for any halos change linearly with increasing $\alpha$ and is found to be negative less than  $\alpha \simeq-25.13$.
It suggests that the scalarization can only occur in regions where the $\alpha$ is sufficiently negative for any halo.

In the example above, the mass of the central BH is fixed at $M_{BH}=5.62 \times 10^6M_{\odot}$, but in reality the central BH could be much heavier. The mass of DM in the halo also varies largely from halos to halos.  Fig.\ref{CDMDMmass} shows the difference of cases between $M_{_\mathrm{DM}}=1.0 \times 10^{10}M_{\odot}$ and $M_{_\mathrm{DM}}=1.0 \times 10^{12}M_{\odot}$.As can be seen, there is almost no change in the value of $S$ even if $M_{_\mathrm{DM}}$ increases. On the other hand, unlike $M_{_\mathrm{DM}}$, $S$ is clearly dependent on the mass $M_{BH}$ of the BH. Fig.\ref{CDMBHmass} shows the difference of cases of $M=1.0 \times 10^6 \sim10^{9}M_{\odot}$. The value of $S$ becomes smaller the bigger the mass $M$ of the BH. 
Intuitively, it would seem that the heavier the BH and the higher the central density of DM, the stronger the scalarization, but in fact the opposite is true. This is because the scalarization  (the scalar hair) has the expanse and depends on the configuration of the material (metric).
Therefore, the heavier the BH, the more the hair is collapsed, and thus scalarization is considered to be conversely weaker.

\begin{figure*}
\centering
\begin{minipage}{1.0\columnwidth}
\centering
\includegraphics[width=\columnwidth]{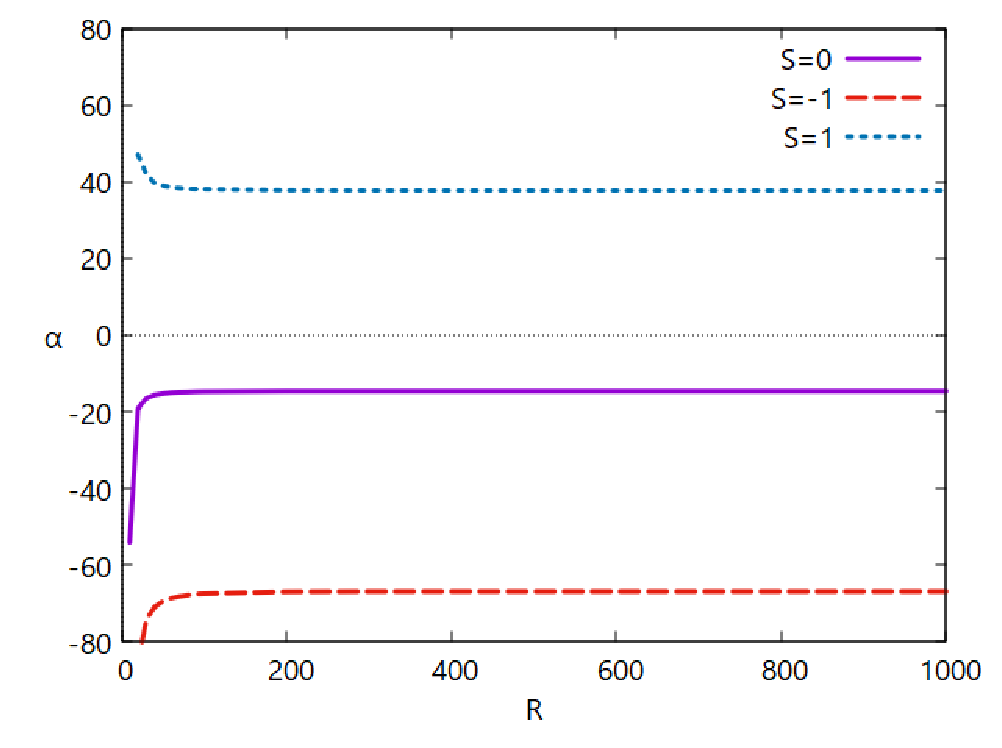}
\caption{\label{SFDMSplane}The $R-\alpha$ plane. Each line in the figure represents $S=-1.0, 0, 1.0$, respectively (SFDM, $1\le R \le1000$, $M_{_\mathrm{DM}}= 1.0, \ell=0$).}
\end{minipage}
\begin{minipage}{1.0\columnwidth}
\centering
\includegraphics[width=\columnwidth]{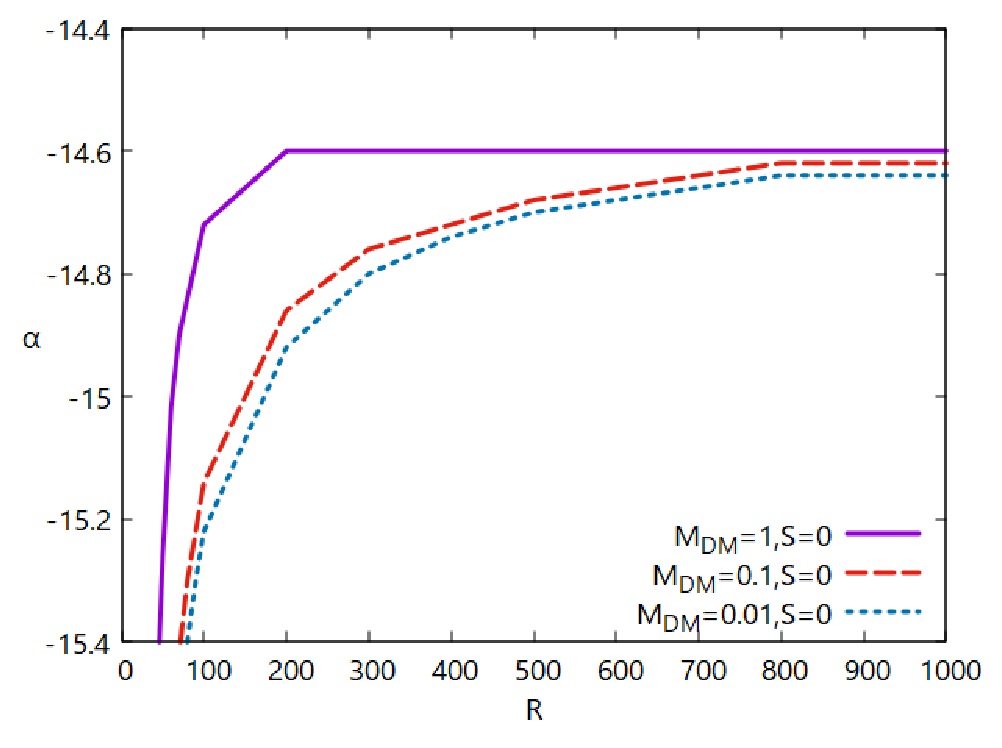}
\caption{\label{SFDMSplaneMDM}The lines of $S=0$ for $M_{_\mathrm{DM}}= 0.01, 0.1$ and $1.0$ on the $R-\alpha$ plane (SFDM,  $1\le R \le1000$, $\ell=0$).}
\end{minipage}
\end{figure*}

\begin{figure*}
\centering
\begin{minipage}{1.0\columnwidth}
\centering
\includegraphics[width=\columnwidth]{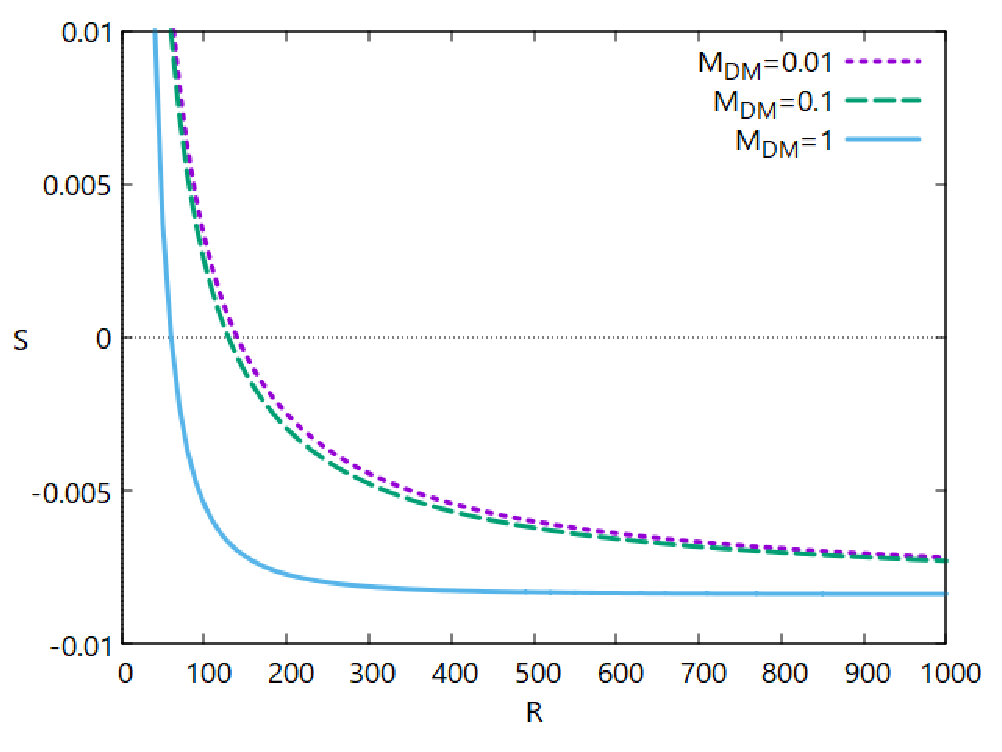}
\caption{\label{SFDMalpha-15}The value of S with changing $R$ from $1.0$ to $1000$ for $M_{_\mathrm{DM}}= 0.01, 0.1$ and $1.0$ and fixing $\alpha=-15$ (SFDM, $\ell=0$).}
\end{minipage}
\begin{minipage}{1.0\columnwidth}
\includegraphics[width=\columnwidth]{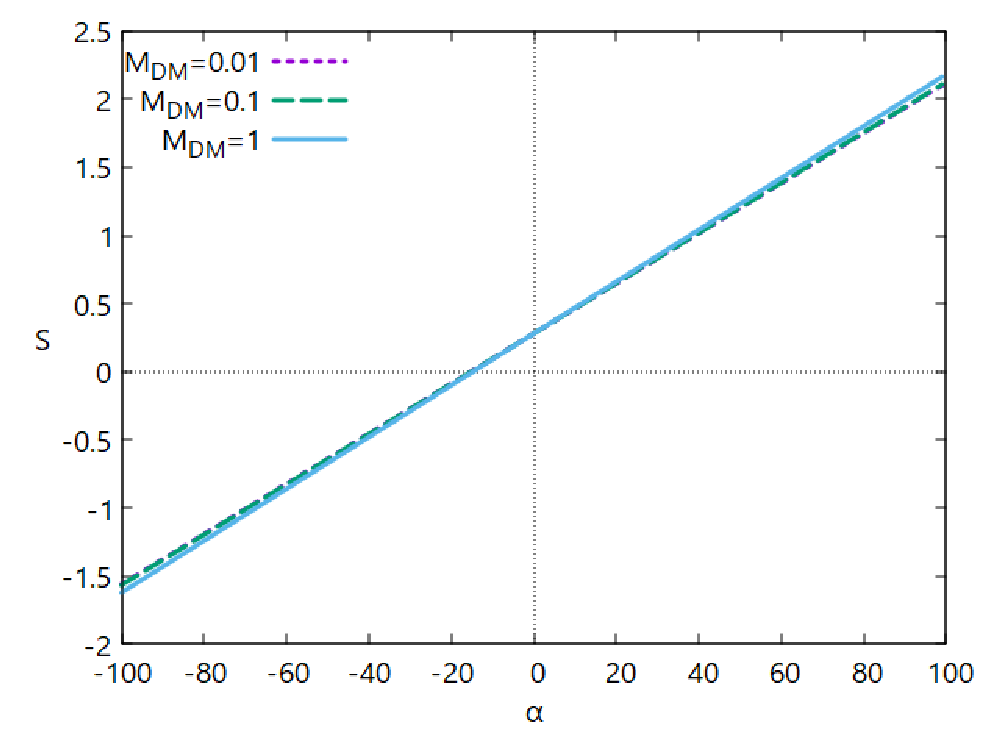}
\caption{\label{SFDMMDM0.01-1R100}The value of S with changing $\alpha$ from $-100$ to $100$ for $M_{_\mathrm{DM}}= 0.01, 0.1$ and $1.0$ and fixing $R=100$ (SFDM, $\ell=0$).}
\end{minipage}
\end{figure*}
\begin{figure*}
\centering
\begin{minipage}{1.0\columnwidth}
\centering
\includegraphics[width=\columnwidth]{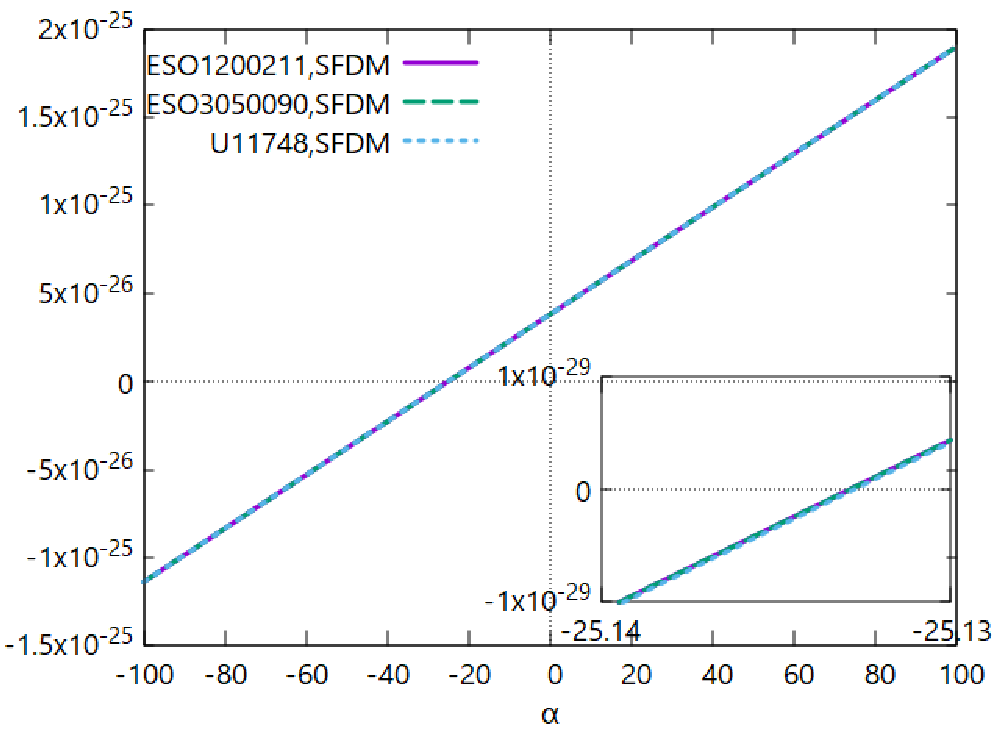}
\caption{\label{sfdmcompare}The value of $S$ for ESO1200211, ESO3050090 and U11748 with changing $\alpha$  (SFDM,$\ell=0$). The inset figure is an enlarged view of the area where each line is S=0.}
\label{ESO1200211sfdm}
\end{minipage}
\begin{minipage}{1.0\columnwidth}
\includegraphics[width=\columnwidth]{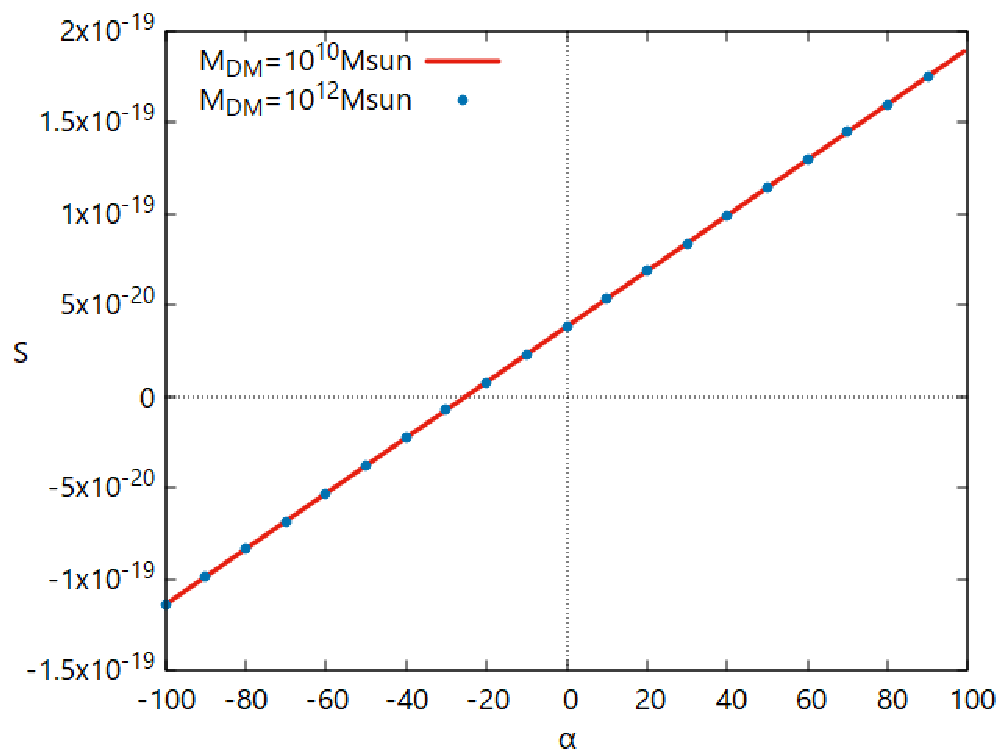}
\caption{\label{SFDMDMmass}The value of $S$ with $M_{_\mathrm{DM}}=1.0 \times 10^{10}M_{\odot}$ and $1.0 \times 10^{12}M_{\odot}$, changing $\alpha$ (SFDM, $\ell=0$,$R=10$kpc, $M=1.0 \times 10^6M_{\odot}$).}
\end{minipage}
\end{figure*}

\subsection{SFDM}
Next we calculate the general case of SFDM. DM halos of SFDM are static Bose-Einstein condensates (BEC). 
This model of the density profile (BEC profile) is \cite{2011JCAP...05..022H}
\begin{equation}
{\rho}_{_\mathrm{SFDM}}=\rho_s\frac{\sin(k r)}{kr} ,
\end{equation}
$\rho_s$ is the central density of the condensates. 
We relate $\rho_s$ and $M_{_\mathrm{DM}}$ from the relation
\begin{align}
M_{_\mathrm{DM}}&=\int_{r_g}^{R_s}\mathrm{d}r4\pi r^2 {\rho}_{_\mathrm{SFDM}} \\
&= \int_{r_g}^{R_s}\mathrm{d}r \frac{4 \pi \rho_s r^2 \sin(kr)}{kr} .
\end{align}
Fig.\ref{SFDMSplane} shows $S=-1.0, 0, 1.0$ for $M_{_\mathrm{DM}}= 1.0$, and fig.\ref{SFDMSplaneMDM} shows $S=0$ for $M_{_\mathrm{DM}}=0.01, 0.1, 1.0$ in the $R-\alpha$ plane. The meaning of the figures is as same as fig.\ref{CDMSplane} and .\ref{CDMSplaneMDM} of CDM, but the behavior of $S$ is different in extremely small $R$ region. CDM has a smaller $S$, while SFDM has a larger S (Fig.\ref{SFDMalpha-15}). In the case of SFDM, scalarization is less likely to occur for a smalll halo.


\begin{figure}
\includegraphics[width=\columnwidth]{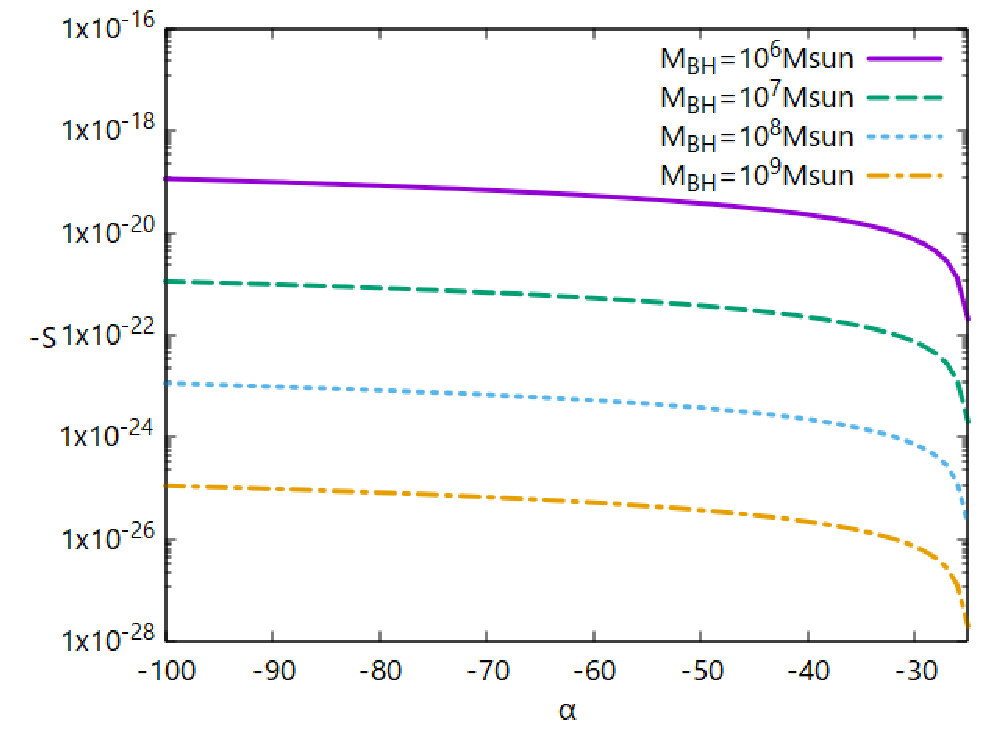}
\caption{\label{SFDMBHmass}The value of $S$ with $M=1.0 \times 10^6M_{\odot} -10^{9}M_{\odot}$, changing $\alpha$ (SFDM, $\ell=0$,$R=10$kpc, $M_{_\mathrm{DM}}=1.0 \times 10^{10}M_{\odot}$).}
\end{figure}

\subsection{SFDM for actual halo}

In this subsection, we will examine the scalarization for specific halos in the case of SFDM. The halos used in the examples are the same as in CDM, and the parameters of the fitting in the case of SFDM are listed in the table \ref{table2}. Fig.\ref{sfdmcompare} show results of (\ref{area})  for ESO1200211, ESO3050090 and U11748. The value of $\alpha$ at which scalarization occurs in the SFDM model is almost the same as the result for the CDM model, with $S$ negative for $\alpha \simeq 25.13$. Fig.\ref{SFDMDMmass} and .\ref{SFDMBHmass} show the dependence of DM mass and BH mass as in the case of SFDM. These results remain unchanged in comparison with  CDM.\par
\begin{figure}
\includegraphics[width=\columnwidth]{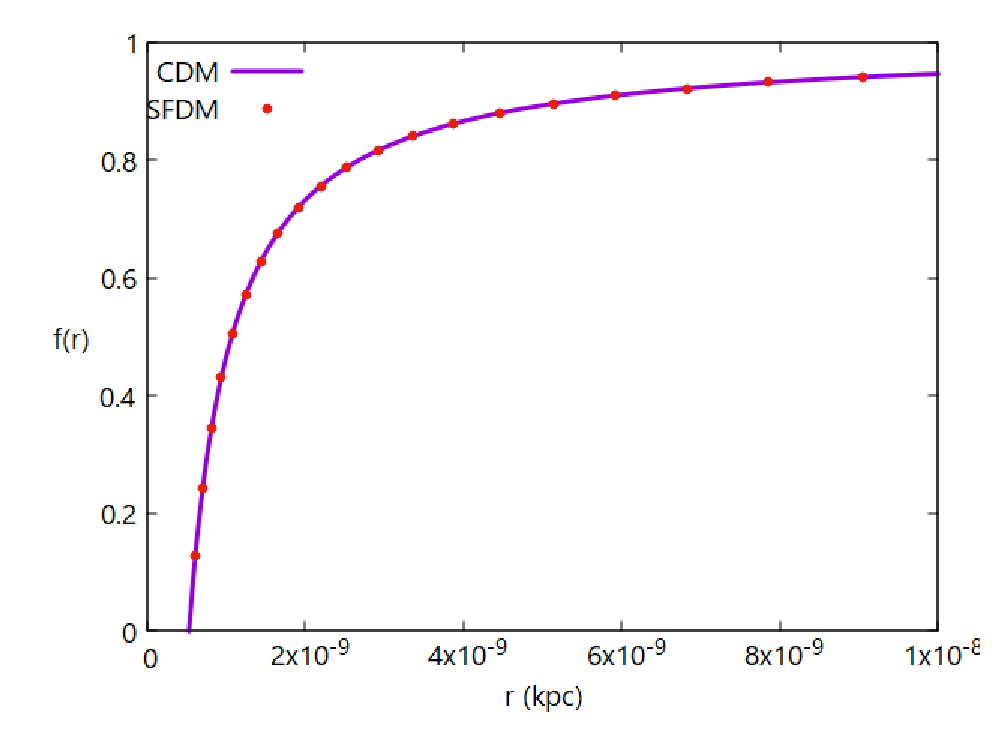}
\caption{\label{cdmsfdmmetric}The value of metric $f(r)$ of CDM and SFDM  for ESO1200211.}
\end{figure}

These indicate that the observation of the scalarization of the actual halo does not distinguish between CDM and SFDM.
Scalarization is mainly determined by the halo metric. The halo metric is determined by the fitting of the galaxy's rotation curve, and there is almost no difference in metrics between CDM and SFDM for the same galaxy (Fig.\ref{cdmsfdmmetric}).

\begin{table}
\caption{table2}
\label{table2}
\begin{ruledtabular}
\begin{tabular}{cccccccc}
Halo &$\rho_s(\times 10^{-3} M_{\odot}pc^{-3})$&$R(kpc)$ \\ \hline
ESO1200211&13.66&2.92 \\
ESO3050090&21.50&4.81 \\
U11748&228.42&7.67 \\ \hline
\end{tabular}
\end{ruledtabular}
\end{table}



\begin{figure*}[t]
\centering
\begin{minipage}{1.0\columnwidth}
\centering
\includegraphics[width=\columnwidth]{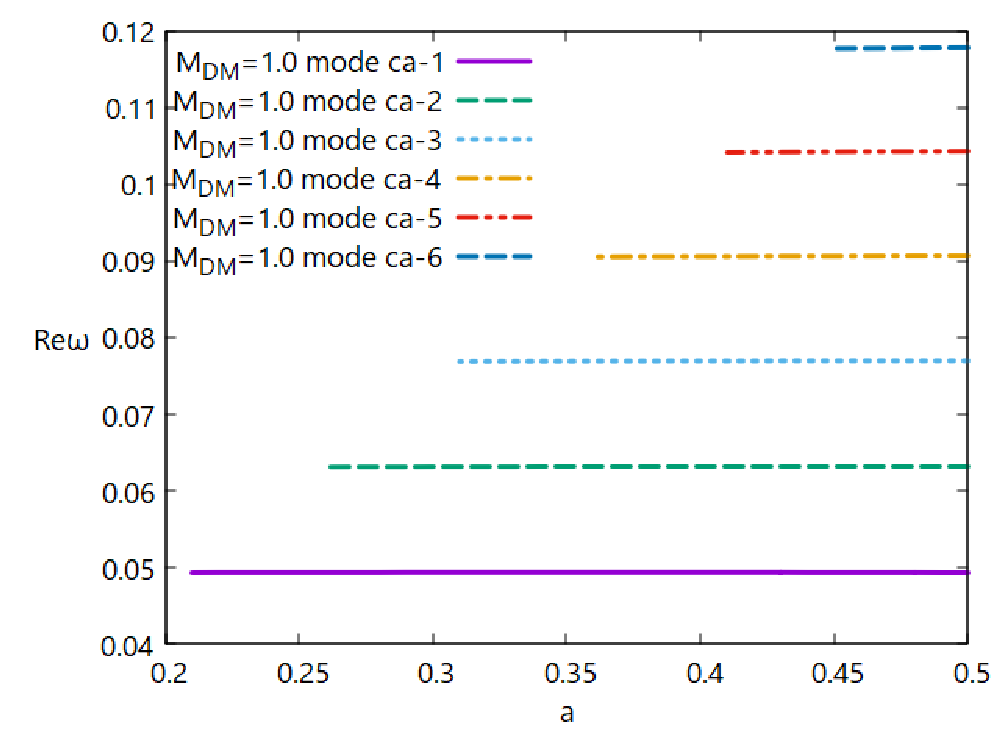}
\caption{Real part of eigenvalue $\omega$ as a function of $a$ (CDM, $\alpha=1.0$, $\ell=1,m=1$, $M_{_\mathrm{DM}}=1.0$).}
\label{cdmaRe}
\end{minipage}
\begin{minipage}{1.0\columnwidth}
\centering
\includegraphics[width=\columnwidth]{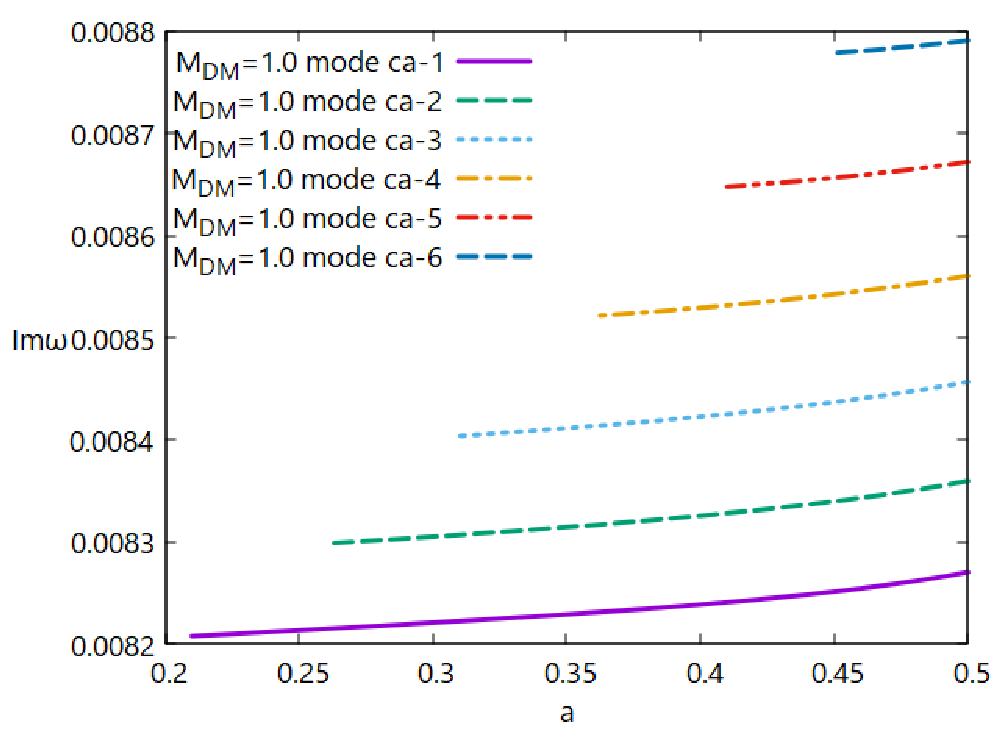} 
\caption{Imginary part of eigenvalue $\omega$ as a function of $a$ (CDM, $\alpha=1.0$, $\ell=1,m=1$, $M_{_\mathrm{DM}}=1.0$)}
\label{cdmaIm}
\end{minipage}
\end{figure*}

\section{The superadiant instability}
\label{SIsec}
In this section, we will look at the superradiant instability. For superradiant instability to occur, a scalar field injected into the rotating BH needs to satisfy the superradinat condition (\ref{superradint condition}). The metric of the rotating BH spacetime surrounded by the DM halo is \cite{2023EPJC...83..565L}
\begin{align}\label{rot metric}
ds^2=&-\left(1-\frac{r^2+2Mr-r^2f(r)}{\Sigma^2}\right)dt^2+\frac{\Sigma^2}{\varDelta}dr^2 \notag \\
&+\Sigma^2 d\theta^2+\frac{A\sin^2\theta}{\Sigma^2}d\varphi^2 \notag \\
&-\frac{2(r^2+2Mr-r^2f(r))a\sin^2\theta}{\Sigma^2}d\varphi dt ,
\end{align}
where
\begin{align}
\varDelta&=r^2f(r)-2Mr+a^2 ,\\
\Sigma^2&=r^2+a^2\cos^2\theta , \\
A&=(r^2+a^2)^2-a^2\varDelta\sin^2\theta .
\end{align}
Here $M$ is the mass of BH, $a$ is the rotational parameter, and $f(r)$ is a part of the metric. Notice that this $f(r)$ is not the same $f(r)$ as in section \ref{DMsec}. In the case of CDM, $f(r)$ is
\begin{equation}
f_c(r)=\left(1+\frac{r}{R_s}\right)^{-{8\pi  \rho_c R_s^3}/{ r}} ,
\end{equation}
and in the case of SFDM,
\begin{equation}
f_s(r)=\exp \left(\frac{-8G\rho_c R^2}{ \pi}\frac{\sin(\pi r/R)}{\pi r/R}\right) .
\end{equation}
The definitions of $\rho_c$ $R_s$, and $R$ are the same as in the above section. Also, for convenience in this section, the unit system $c=G=M=r_g=1$ is used. ($r_g=GM/c^2$). \par 
The Klein-Gordon (KG) equation in the spacetime of rotating BH with DM halo (\ref{rot metric}) is

\begin{align}
&-\frac{A}{\varDelta \Sigma^2}\partial^2_t \Psi+\frac{2a(-2Mr-r^2+r^2f(r))}{\varDelta \Sigma^2}\partial_t \partial_{\phi}\Psi \notag \\
&+\frac{1}{\Sigma^2 \sin \theta}\partial_{\theta}(\sin \theta \partial_{\theta} \Psi)+\frac{1}{\Sigma^2}\partial_r (\varDelta \partial_r \Psi)\notag \\
&+\frac{-2Mr-r^2+\Sigma^2+r^2 f(r)}{\varDelta \Sigma^2 \sin^2 \theta}\partial_{\phi}^2 \Psi=\mu^2(r,\theta) \Psi .
\end{align}
To investigate the superradiant instability of a rotating BH, it is necessary to solve the KG equation. Usually, separation of variables is performed, the equations are separated in the radial and angular directions, and the respective eigenvalue equations are solved. However, when considering the massive scalar, the mass $\mu_0$ on the right-hand side of KG equation is a constant, but not when considering DM-derived effective mass. The effective mass (\ref{effective mass}) can be calculated from the rotational BH metric (\ref{rot metric}), but it is a very complicated polynomial of $r$ and $\theta$, and in general $r$ and $\theta$ directions of KG equations are not separable variables. However, if we take the slow rotation approximation, we can divide the effective mass into the part that depends only on $r$ and the other parts and obtain the following equation:

\begin{figure*}[t]
\centering
\begin{minipage}{1.0\columnwidth}
\centering
\includegraphics[scale=0.5]{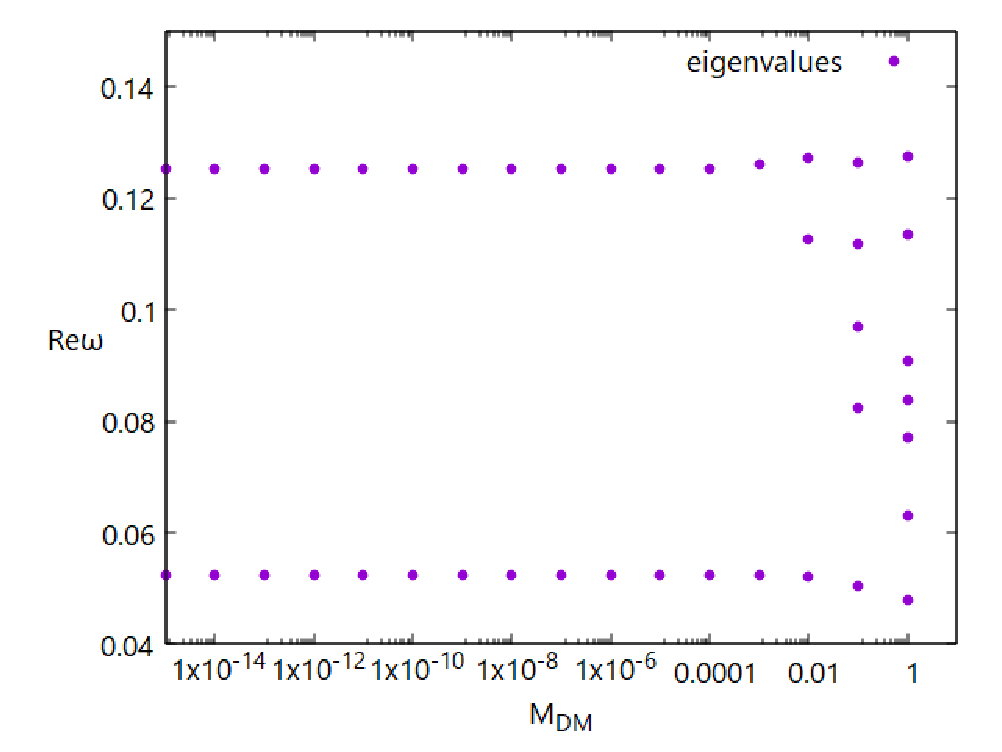}
\caption{Real part of eigenvalue $\omega$ as a function of  $M_{_\mathrm{DM}}$(CDM, $\ell=1,m=1, a=0.5, \alpha=1$).}
\label{cdmMDMRe}
\end{minipage}
\begin{minipage}{1.0\columnwidth}
\centering
\includegraphics[scale=0.5]{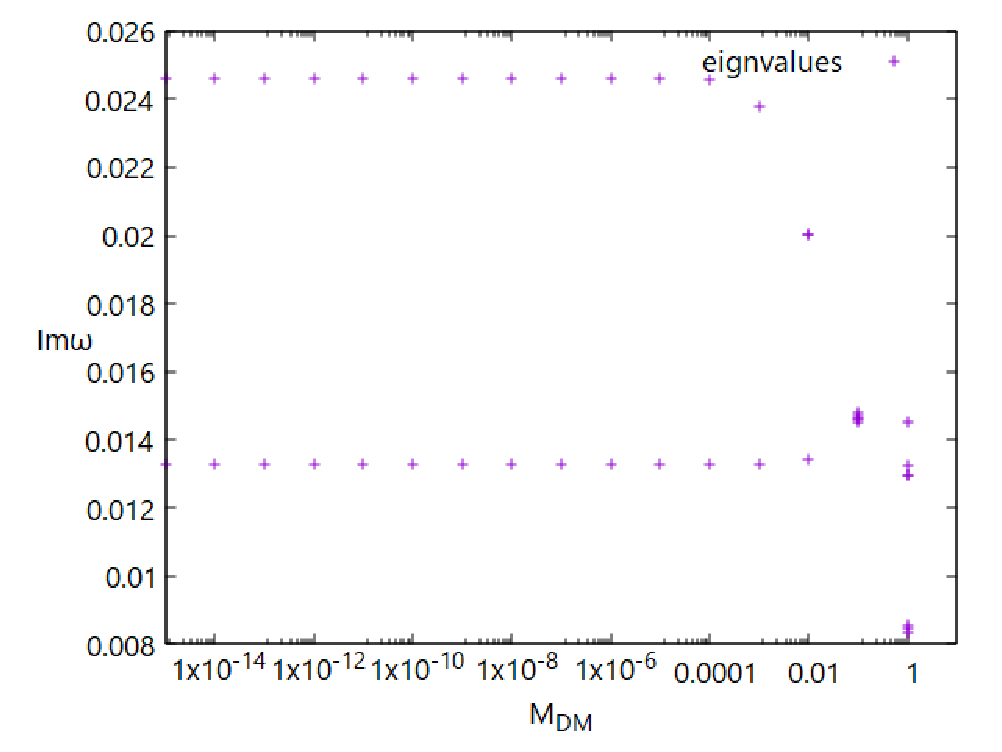}
\caption{Imginary part of eigenvalue $\omega$ as a function of $M_{_\mathrm{DM}}$(CDM, $\ell=1, m=1, a=0.5, \alpha=1$).}
\label{cdmMDMIm}
\end{minipage}
\end{figure*}

\begin{equation}\label{low speed effective}
\mu_{eff}^2(r,\theta)=\frac{\alpha}{4\pi} \frac{2-\varDelta_{.rr}}{r^2}+O(a^2) .
\end{equation}
where $\alpha$ is the same as (\ref{coupling}) and $\varDelta_{,rr}$ is the second-order derivative of $\varDelta$ by $r$.
Considering the slow rotation limit, all $a^2$ terms can be ignored. In the following, we will adopt the slow rotation approximation  and will not consider the rapidly rotating state $(a \simeq 1)$ of BHs. (See Appendix.\ref{effmass} for the specific derivation of Eq.(\ref{low speed effective})\par 
Since the first term of the slow rotation approximation (eq.(\ref{low speed effective})) is a function of $r$ only, it can be expressed as $\mu^2(r,\theta)\simeq \mu^2(r)$, allowing for separation of variable. Assuming,
\begin{equation}
\Psi(t,r,\theta,\phi)=e^{-i \omega t}e^{i m \phi}R(r)S(\theta)
\end{equation}
we get the equation (see Apendix.\ref{derivationKG}):
\begin{align}
\frac{1}{\sin\theta}&\partial_{\theta}(\sin\theta \partial_{\theta})S(\theta)\notag \\
&+\left(\Lambda_{\ell m}+a^2\omega^2\cos^2\theta-\frac{m^2}{\sin^2\theta} \right)S(\theta) =0 .
\end{align}
Transforming variable  $x=\cos\theta$, we have,
\begin{align}\label{angular}
&(1-x^2) \frac{d^2 S(x)}{dx^2}-2x\frac{dS(x)}{dx}\notag \\
&+\left(\Lambda_{\ell m}+a^2\omega^2 x^2-\frac{m^2}{1-x^2} \right)S(x)=0 .
\end{align}
The radial equation becomes
\begin{equation}\label{radial}
\varDelta \frac{ d}{dr}\left(\varDelta\frac{ d}{dr} \right)R(r)+[K^2(r)-(\lambda+\mu^2(r)r^2)\varDelta]R(r)=0 .
\end{equation}
Here,
$K^2(r)=\omega^2(r^2+a^2)^2-2am\omega(r^2+a^2)+a^2 m^2$ and $\lambda=\Lambda_{\ell m}+a^2 m^2-2am\omega$, where $\Lambda_{\ell m}$ is the seperation constant. $\omega$ is the frequency of the system to be obtained and in general is a complex number ($\omega=\omega_R+i \omega_I$).
$\omega_I>0$ corresponds to an unstable mode and $ \omega_I<0$ corresponds to a stable mode.
The conditions for superradiance instability are $ \omega_I>0$ and $0<\omega_R<{m a}/{(r_+^2+a^2)}$. \par

The boundary conditions of angular equation are \cite{2023EPJC...83..565L}
\begin{equation}\label{angularcondition}
\lim_{x \to -1} S(x) \sim (1+x)^{|m|/2} ,\quad \lim_{x \to +1} S(x) \sim (1-x)^{|m|/2} .
\end{equation}
However, the boundary conditions of the radial equation is not the same as in \cite{2023EPJC...83..565L}. Generally, the boundary conditons at infinity is expressed as
\begin{equation}
R(r) \sim Be^{-k r}+Ce^{+k r} .
\end{equation}
where $k_{\infty}=\sqrt{\mu_{\infty}^2-\omega^2}$ and $\mu_{\infty}=\lim_{r \to \infty} \mu_{\infty}=\mu_{\mathrm{eff}}$.

As discussed in \cite{2022PhRvD.106b4007L}, for the massive scalar boson in GR, $C=0$ corresponds to a exponentially damping solution at infinity due to the asymptotic mass. On the other hand, in the scalar-tensor theory, the potential barrier created by the effective mass exists near BH. Moreover, this potential vanishes at infinity, because the extent of matter is finite. Therefore, the confinement also happens near BH. In this case, the $C=0$ is not a damping solution, but corresponds to an incident wave from infinity. It means an energy injection from infinity, which is not physical, so we must choose $B=0$. It corresponds to outgoing waves at infinity.


The boundary condition at infinity should be
\begin{equation}\label{infinitycondition}
\lim_{x \to \infty}R(r) \sim exp(k r) r^{\beta-1} ,\quad \beta=\frac{(r_++r_-)(\mu(\infty)-2 \omega^2)}{2 k} .
\end{equation}
Here, $\mu(\infty)$ is $0$. Because the scale of the DM halo is finite, so the effective mass vanishes at infinity. Note that the boundary condition on the horizon is the same as usual, considering only incident waves to the horizon. Therefore, we set it as
\begin{equation}
\lim_{x \to r_+}R(r) \sim (r-r_+)^{-i \alpha} ,\quad \alpha=\frac{r_+^2+a^2}{r_+-r_-}(\omega-m \varOmega) .
\end{equation}
\par
\cite{2023EPJC...83..565L} computed $\Lambda_{\ell m}$ and $\omega$ using the continued fraction method.
We will take the same approach for the angular  equation and compute $\Lambda_{\ell m}$. From (\ref{angularcondition}), the solution $S(x)$ of the angular equation (\ref{angular}) is considered to be in the form of a continuous solution like :
\begin{equation}
S(x)=\exp(akx)(1-x)^{|m|/2}(1+x)^{|m|/2}\MM^{\infty}_{n=0} b_n (1+x)^n .
\end{equation}
Substituting this into (\ref{angular}), we obtain an incremental equation for $b_n$:
\begin{equation}\label{recurrence relation}
\begin{cases}
\alpha_0 b_1+\beta_0 b_0=0 ,\\
\alpha_n b_{n+1}+\beta_n b_n+\gamma_n b_{n-1}=0 .
\end{cases}
\end{equation}
Here, $b_0=1$, each coefficiets are
\begin{equation}
\begin{cases}
\alpha_n=-2(n+1)(|m|+n+1) ,\\
\beta_n=-\Lambda_{lm}+|m|(-2ak+2n+1)-ak(ak+2) ,\\
\quad \quad+m^2+n^2-4akn+n ,\\
\gamma_n=2ak(|m|+n) .
\end{cases}
\end{equation}
Transforming the recurrence relation (\ref{recurrence relation}),
\begin{equation}
\frac{b_n}{b_{n-1}}=-\frac{\gamma_n}{\beta_n+\alpha_n b_{n+1}/b_n} .
\end{equation}
Using this relation repeatedly from $n=0$, we get
\begin{equation}
0=\beta_0-\frac{\alpha_0 \gamma_1}{\beta_1-\frac{\alpha_1 \gamma_2}{\beta_2-\frac{\alpha_2 \gamma_3}{\beta_3-...}}} .
\end{equation}

This equation is the continued fraction to be solved. The procedure of the continued fraction method is to obtain $\Lambda_{lm}$ by solving this continued fraction equation numerically using the (complex) Newton method with $a$, $k$, and $m$ given. We truncate terms higher than $n=1000$.\par
The same method cannot be used for the Radial equation because it contains an effective mass $\mu^2(r)$ that is not a constant. Therefore, in this study, the solution of the angular equation $\Lambda_{\ell m}$ is obtained by the continued fraction method \cite{2006PhRvD..73l4040K,2007PhRvD..76h4001D,Leaver:1985ax,2023EPJC...83..565L}, and the solution of the radial equation is calculated by the Direct integration method (Shooting method) \cite{2010PhRvD..81l4021M}. Direct integration method integrates the radial equation both from the (outer) horizon and the infinity, and matches the two solutions at the matching point $r_m$. In this case,  the Muller method \cite{10.5555/573178} is used to search for eigenvalues by finding roots of Wronskian.  In this method, first, three candidate eigenvalues are prepared, calculated by the Shooting method, and the Wronskian is obtained. Next, the same calculation is repeated using the new candidate eigenvalues obtained by the Muller method, until the residual beomes less than or equal to a given tolerance.
\par

\begin{figure*}[t]
\centering
\begin{minipage}{1.0\columnwidth}
\centering
\includegraphics[scale=0.5]{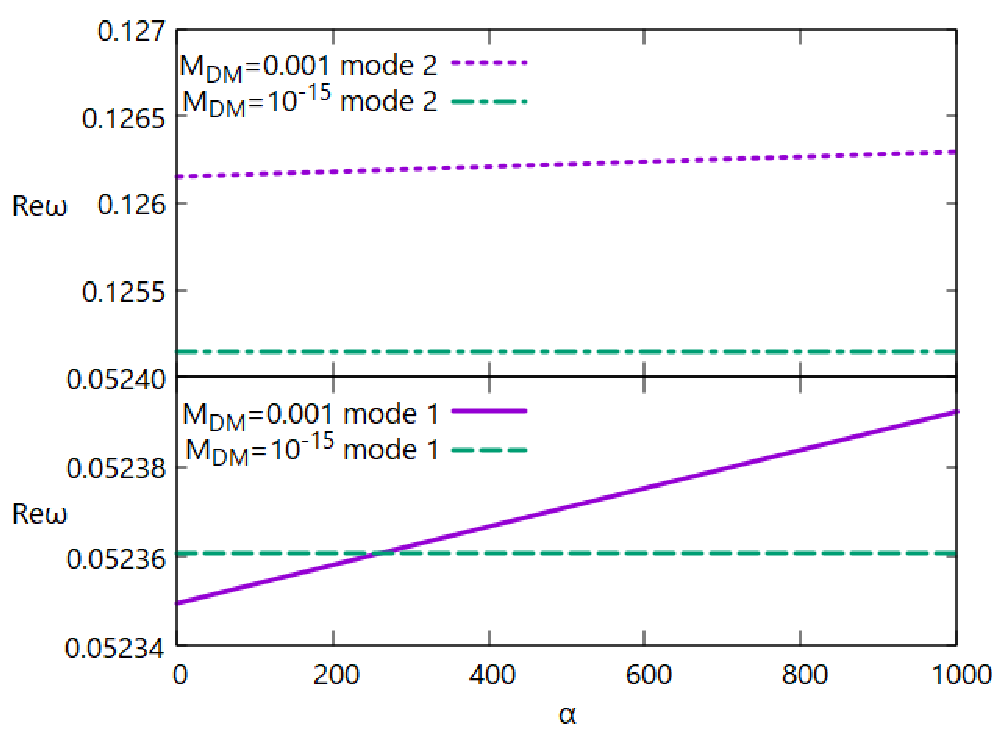}
\caption{Real part of eigenvalue $\omega$ as a function of  $\alpha$ for $M_{_\mathrm{DM}}=0.001$ and $10^{-15}$ (CDM,  $\ell=1,m=1, a=0.5$).}
\label{cdmalphaRe}
\end{minipage}
\begin{minipage}{1.0\columnwidth}
\centering
\includegraphics[scale=0.5]{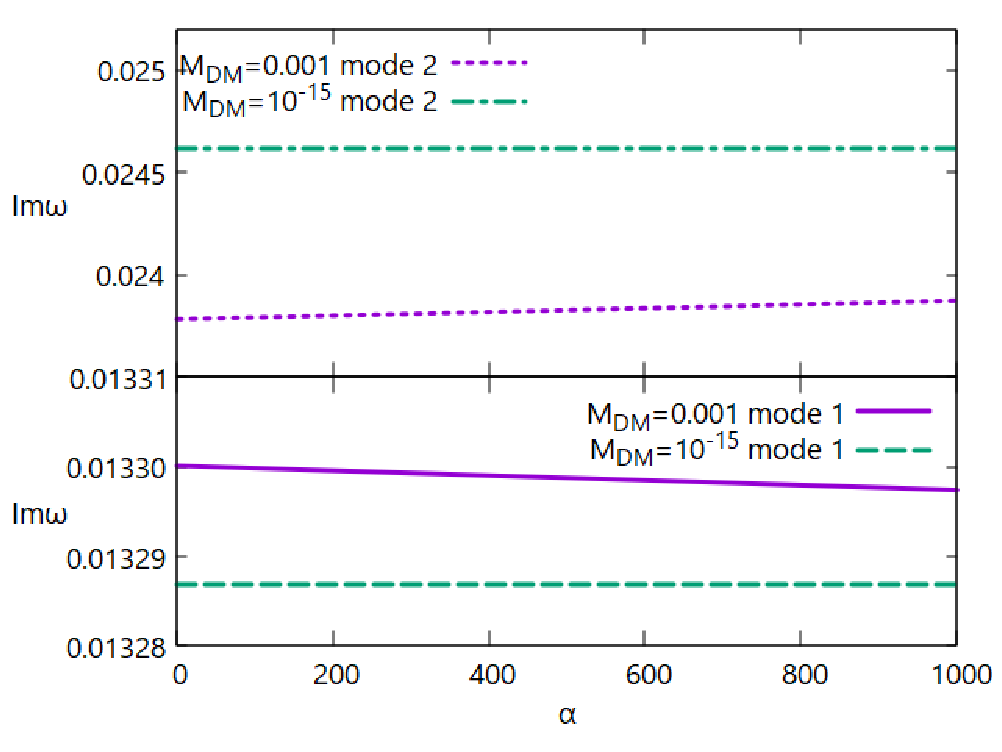}
\caption{Imginary part of eigenvalue $\omega$ as a function of $\alpha$ for $M_{_\mathrm{DM}}=0.001$ and $10^{-15}$ (CDM, $\ell=1,m=1, a=0.5$).}
\label{cdmalphaIm}
\end{minipage}
\end{figure*}

\subsection{CDM}
We first show the CDM model. 
The derivatives of $f_c(r)$:
\begin{align}
f_c'(r)=&\left(1+\frac{r}{R_s}\right)^{-{8\pi \rho_c R_s^3}/{ r}}\notag \\
 &\times\left[\frac{8\pi  \rho_c R_s^3}{ r^2} \log\Big(1+\frac{r}{R_s}\Big) -\frac{8\pi  \rho_c R_s^3}{ r(r+R_s)}  \right] ,
\end{align}
\begin{align}
f_c''(r)&=\left(1+\frac{r}{R_s}\right)^{-{8\pi \rho_c R_s^3}/{ r}}\Big[-\frac{16\pi  \rho_c R_s^3}{ r^3}\log\Big(1+\frac{r}{R_s}\Big) \notag \\
&+ \frac{8\pi  \rho_c R_s^3}{r^2(r+R_s)}-\frac{8\pi  \rho_c R_s^3}{ r}\frac{1}{R_s}\Big (\frac{1}{(r+R_s)^2}-\frac{1}{r^2}\Big)\Big] \notag \\
&+f'_c(r)\times\Big[\frac{8\pi  \rho_c R_s^3}{ r} \log\Big(1+\frac{r}{R_s}\Big)\notag -\frac{8\pi  \rho_c R_s^3}{ r(r+R_s)} \Big] ,
\end{align}
Here, ' is the differentiation with respect to $r$, and
\begin{align}
\varDelta=r^2f(r)-2Mr+a^2 ,\\
\varDelta''=r^2 f''(r)+4rf'(r)+2f(r) .
\end{align}
We then caluculate the effective mass (\ref{low speed effective}) as,
\begin{align}\label{effective mass2}
\mu_{eff}^2(r,\theta)&\simeq \frac{\alpha}{4\pi} \frac{2-\varDelta^{''}}{r^2}\notag \\
&=\frac{\alpha}{4\pi} \frac{2-(r^2 f''(r)+4rf'(r)+2f(r))}{r^2} .
\end{align}
\par

The angular equation (\ref{angular}) and the radial equation (\ref{radial}) are solved for the eigenvalue $\omega$. Fig.\ref{cdmaRe} and \ref{cdmaIm} show Re[$\omega$] and Im[$\omega$] as a function of Kerr parameter $a$ with $R_s=100,\alpha=1,\ell=1,m=1$. The number of modes varies with $a$ and  increases as $a$ is increased. There is a mode (mode ca-1) with the smallest Im[$\omega$], and as $a$ increases, multiple modes (mode ca-2 - ca-6) appear in addition to mode ca-1. On the other hand, no mode appeared for $a$ smaller than $a \simeq 0.2$. This indicates that superradiant instability does not appear when the rotation is too slow.
\par

\begin{figure*}[ht]
\begin{minipage}{1.0\columnwidth}
\centering
\includegraphics[scale=0.5]{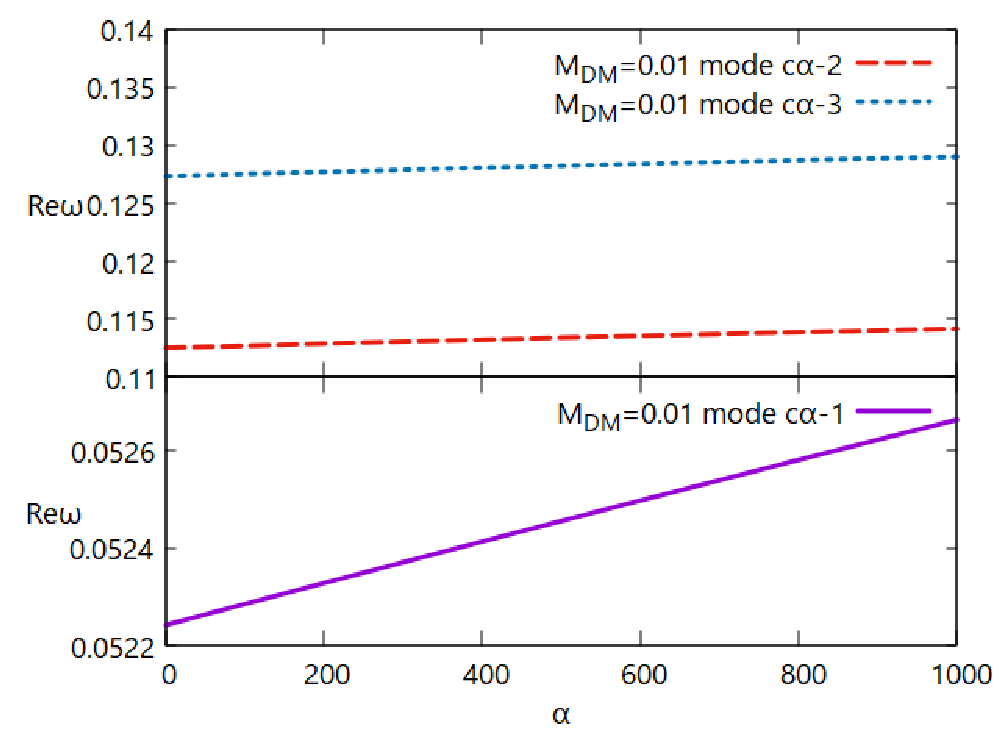}
\caption{Real part of eigenvalue $\omega$ as a function of  $\alpha$(CDM, $\ell=1,m=1, $a$=0.5, M_{_\mathrm{DM}}=0.01$).}
\label{cdmalphaMDM0.01Re}
\end{minipage}
\begin{minipage}{1.0\columnwidth}
\centering
\includegraphics[scale=0.5]{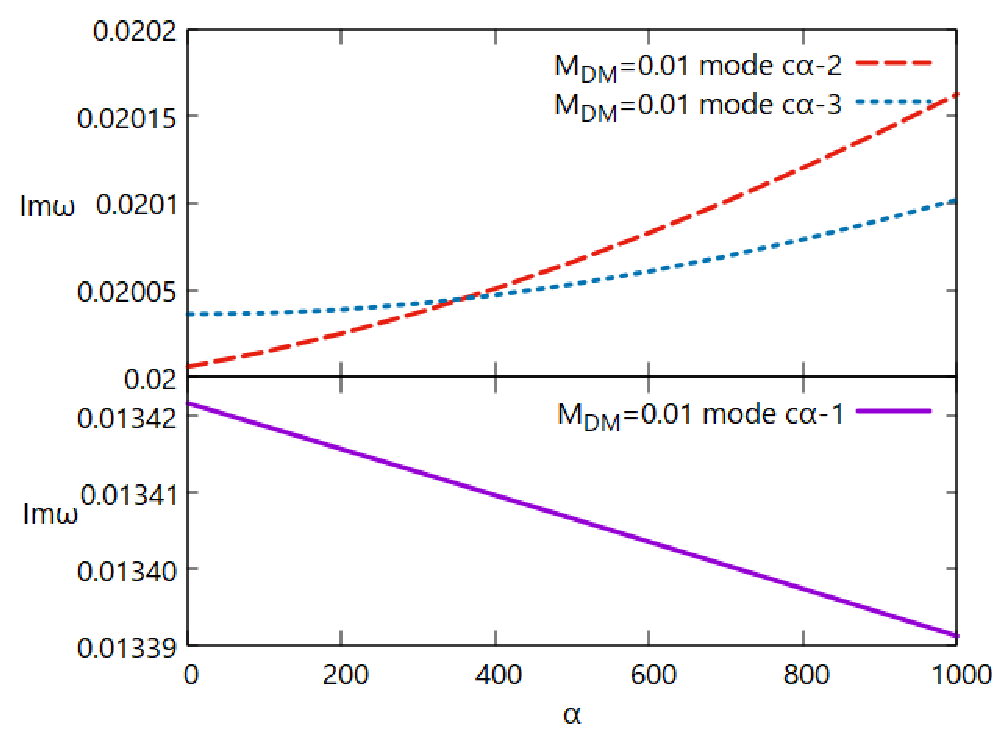}
\caption{Imginary part of eigenvalue $\omega$ as a function of $\alpha$(CDM,$\ell=1,m=1,a=0.5,M_{_\mathrm{DM}}=0.01$).}
\label{cdmalphaMDM0.01Im}
\end{minipage}
\end{figure*}

The number of modes and how they change are affected by $M_{_\mathrm{DM}}$. It is limited to halo masses above a certain weight. Fig.\ref{cdmMDMRe} and \ref{cdmMDMIm} show Re[$\omega$] and Im[$\omega$] at $a=0.5$ , $R_s=100$ for each $M_{_\mathrm{DM}}$. For $M_{_\mathrm{DM}}$ smaller than $M_{_\mathrm{DM}}=0.01$, the number of modes is only two and also shows almost no change due to $M_{_\mathrm{DM}}$. However, as $M_{_\mathrm{DM}}$ becomes larger than $0.01$, the number of eigenvalues increases. Furthermore, for larger $M_{_\mathrm{DM}}$, Im[$\omega$] becomes smaller, thus the stability of the system itself increases as the mass of DM increases.\par


\begin{figure*}
\centering
\begin{minipage}{1.0\columnwidth}
\centering
\includegraphics[scale=0.5]{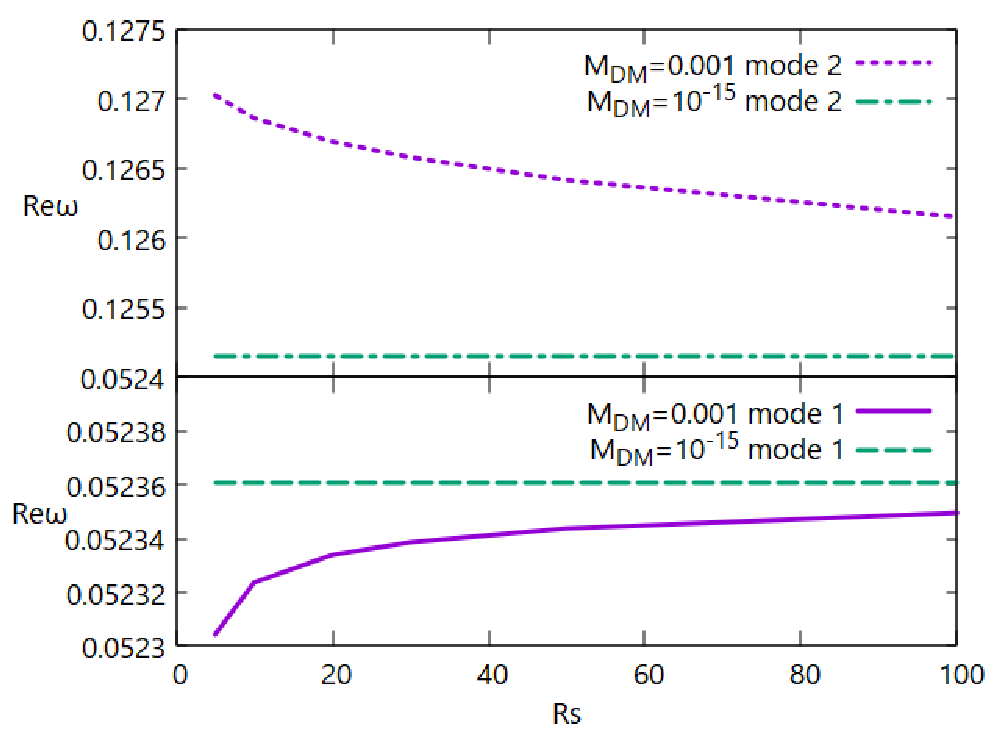}
\caption{Real part of eigenvalue $\omega$ as a function of $R_s$ for $M_{_\mathrm{DM}}=10^{-15}, 0.001$ (CDM, $\alpha=1, \ell=1, m=1, a=0.5$}
\label{cdmRsRe}
\end{minipage}
\begin{minipage}{1.0\columnwidth}
\centering
\includegraphics[scale=0.5]{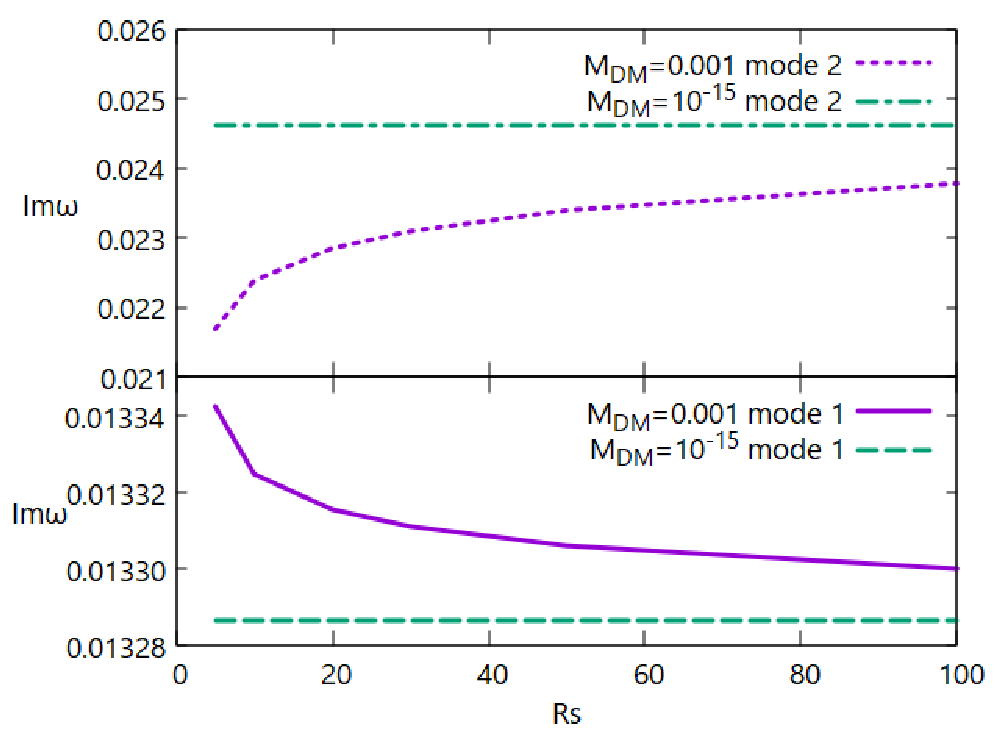}
\caption{Imginary part of eigenvalue $\omega$ as a function of $R_s$ for $M_{_\mathrm{DM}}=10^{-15}, 0.001$ (CDM, $\alpha=1, \ell=1, m=1, a=0.5$}
\label{cdmRsIm}
\end{minipage}
\end{figure*}

\begin{figure*}
\centering
\begin{minipage}{1.0\columnwidth}
\centering
\includegraphics[scale=0.5]{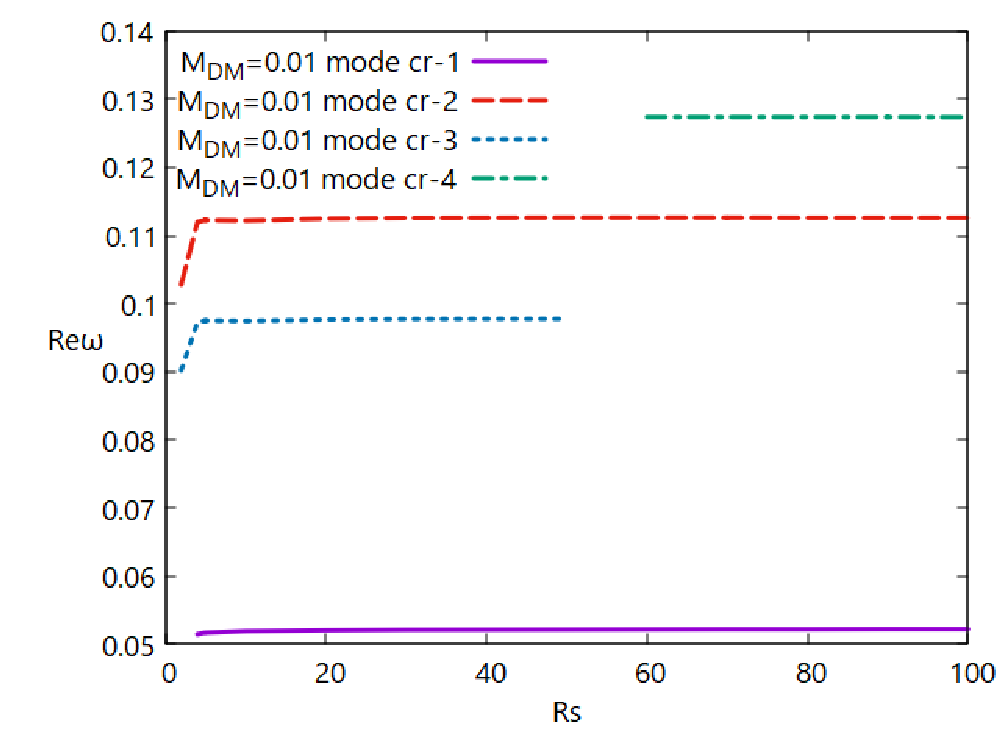}
\caption{Real part of eigenvalue $\omega$ as a function of  $R_s$ for $M_{_\mathrm{DM}}=0.01$(CDM, $\alpha=1, \ell=1, m=1, a=0.5$}
\label{cdmRsMDM0.01Re}
\end{minipage}
\begin{minipage}{1.0\columnwidth}
\centering
\includegraphics[scale=0.5]{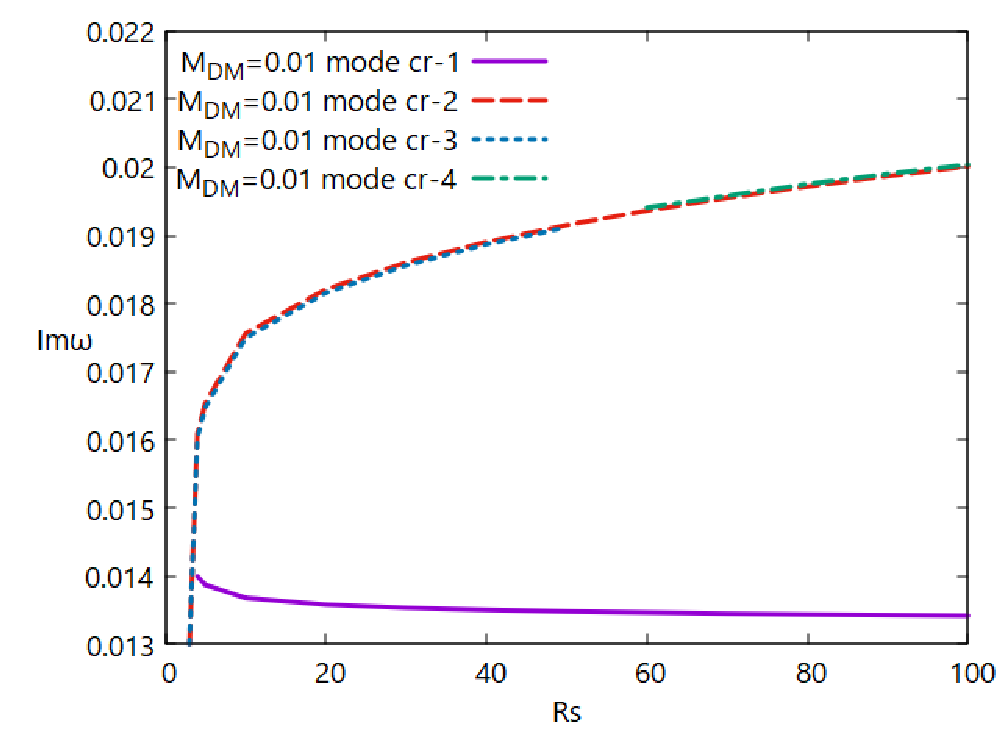}
\caption{Imginary part of eigenvalue $\omega$ as a function of $R_s$ for $M_{_\mathrm{DM}}=0.01$(CDM, $\alpha=1, \ell=1, m=1, a=0.5$}
\label{cdmRsMDM0.01Im}
\end{minipage}
\end{figure*}

\begin{figure*}
\centering
\begin{minipage}{1.0\columnwidth}
\centering
\includegraphics[scale=0.5]{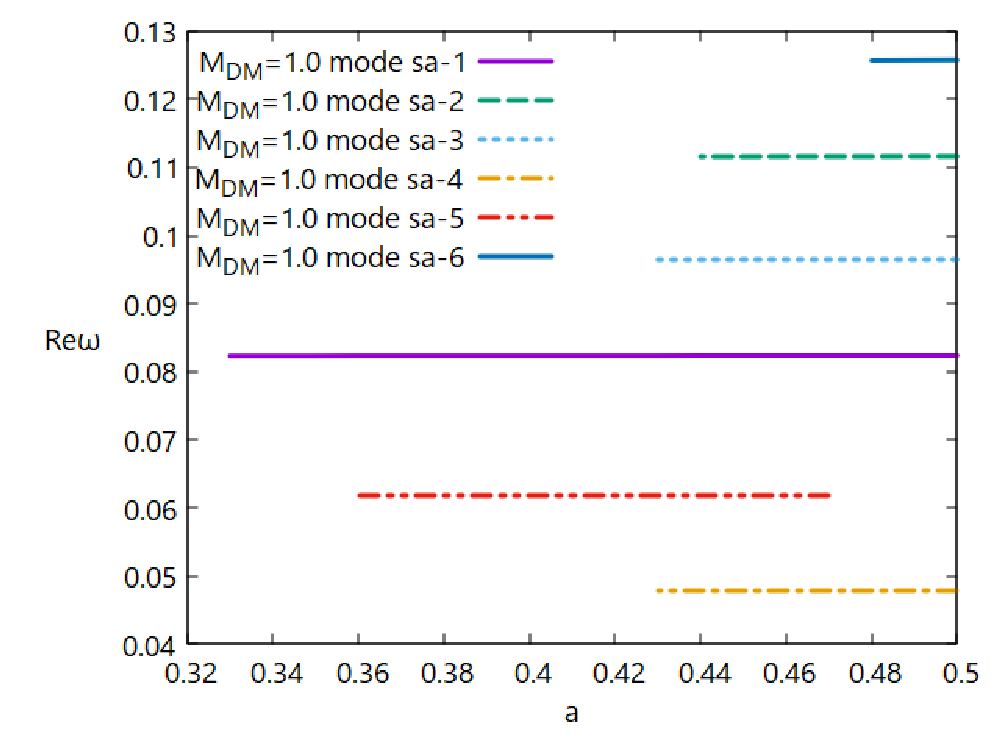}
\caption{Real part of eigenvalue $\omega$ as a function of $a$ (SFDM, $\ell=1, m=1$, $M_{_\mathrm{DM}}=1.0$).}
\label{sfdmaRe}
\end{minipage}
\begin{minipage}{1.0\columnwidth}
\centering
\includegraphics[scale=0.5]{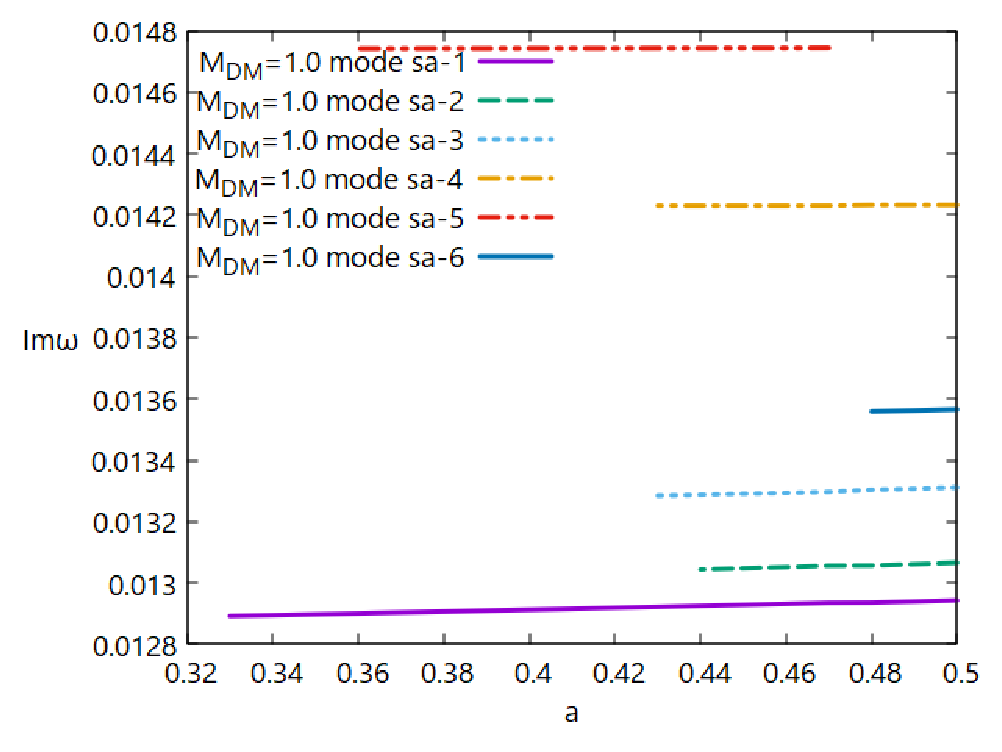} 
\caption{Imginary part of eigenvalue $\omega$ as a function of $a$ (SFDM, $\ell=1, m=1$, $M_{_\mathrm{DM}}=1.0$).}
\label{sfdmaIm}
\end{minipage}
\end{figure*}


Fig.\ref{cdmalphaRe}$-$\ref{cdmalphaMDM0.01Im} shows the Re[$\omega$] and Im[$\omega$] at $a=0.5$ and $R_s=100$ for each $M_{_\mathrm{DM}}=10^{-15}, 0.001, 0.1$. 
When $M_{_\mathrm{DM}}$ is small, there is little change when $\alpha$ is increased. However, as the $M_{_\mathrm{DM}}$ increases, both the real and imaginary parts of the eigenvalues change as the $\alpha$ changes. Furthermore, as mentioned above, the number of modes increases as the $M_{_\mathrm{DM}}$ increases to some extent (Fig. \ref{cdmalphaMDM0.01Im}).
The change of each mode with the increase of $\alpha$ is different for each mode. However, we could not confirm that the number of modes itself increases or decreases with increasing alpha.
Since the magnitude of $\alpha$ affects the height of the potential created by the effective mass, in other words, the height of the potential does not affect the number of modes, but only their amplitudes.
\par



Fig.\ref{cdmRsRe} - \ref{cdmRsMDM0.01Im} show Re[$\omega$] and Im[$\omega$] with changing $R_s$ at $a=0.5$ for $M_{_\mathrm{DM}}=10^{-15}, 0.001,0.01$.
Similarly here, neither amplitude change nor mode increase occurs until $M_{_\mathrm{DM}}$ grows to some extent. However, if $M_{_\mathrm{DM}} = 0.01$ or greater, we can confirm that the number of modes changes with changes in $R_s$ (Fig.\ref{cdmRsMDM0.01Re})

Fig.\ref{cdmRsMDM0.01Re} and .\ref{cdmRsMDM0.01Im} show that the order of magnitude of Re[$\omega$] does not change much, but that Im[$\omega$] is affected by $M_{_\mathrm{DM}}$. As $R_s$ increases, the instability is expected to decrease because the central density $\rho_c$ decreases. As mentioned in the chapter on scalarization, since $T \propto -\rho$ for normal materials, it would seem that the instability also increases as $M_{_\mathrm{DM}}\propto \rho$ increases. \cite{2023EPJC...83..565L} shows that in GR, Im[$\omega$] in the quasi bound state (QBS) decreases as $\rho$ in the DM halo increases. However, in the case of DM halos this is not clearly asserted to be such a characteristic. This is because there are both modes in which Im[$\omega$] increases and decreases with increasing $R_s$.
This is thought to reflect the property that the effective mass of DM is determined by the size of the halo as well as the central density. In other words, the overall structure of the halo, not just the local information around the central BH, affects the strength of the instability.

\subsection{SFDM}
In the case of SFDM metric, the derivatives of $f_s(r)$ are
\begin{align}
f_s'(r)=&-\frac{8\rho_c R^2}{\pi}\frac{\pi}{R} \left[\frac{\cos(\pi r/R)}{\pi r/R}-\frac{\sin(\pi r/R)}{(\pi r/R)^2} \right]\notag \\
&\times\exp \left(-\frac{8\rho_c R^2}{\pi}\frac{\sin(\pi r /R)}{\pi r/R} \right) ,
\end{align}
\begin{align}
f_s''(r)&=\left(-\frac{8\rho_c R^2}{\pi}\right)^2 \left[\frac{\cos(\pi r/R)}{\pi r/R}-\frac{\sin(\pi r/R)}{(\pi r/R)^2} \right]^2\notag \\
&\times\exp \left(-\frac{8G\rho_c R^2}{\pi}\frac{\sin(\pi r /R)}{\pi r/R} \right)\nonumber 
\end{align}
\begin{align}
&-\frac{8\rho_c R^2}{\pi}\Big(\frac{\pi}{R} \Big)^2 \Big[ -\frac{\sin(\pi r/R)}{\pi r/R}-2\frac{\cos(\pi r/R)}{(\pi r/R)^2}\notag \\
&-2\frac{\sin(\pi r/R)}{(\pi r/R)^3} \Big]\notag \\
&\times\exp \left(-\frac{8\rho_c R^2}{\pi}\frac{\sin(\pi r /R)}{\pi r/R} \right) .
\end{align}
The effective mass is (\ref{effective mass2}).\par

For the SFDM case, we also calculated the angular equation (\ref{angular}) and the radial equation (\ref{radial}) to investigate the superradiance instability. Fig.\ref{sfdmaRe} and \ref{sfdmaIm} show the Re[$\omega$] and Im[$\omega$] values for different $a$. As a difference from CDM, the minimum $a$ at which the mode exists is about $0.21$ in CDM, whereas in SFDM the superradiant instability mode exists only at $a\geq 0.33$ or higher. However, Re[$\omega$] is of similar value, while Im$\omega$ is larger for SFDM. SFDM is more unstable than CDM for higher rotation.
\par
\begin{figure*}
\centering
\begin{minipage}{1.0\columnwidth}
\centering
\includegraphics[scale=0.5]{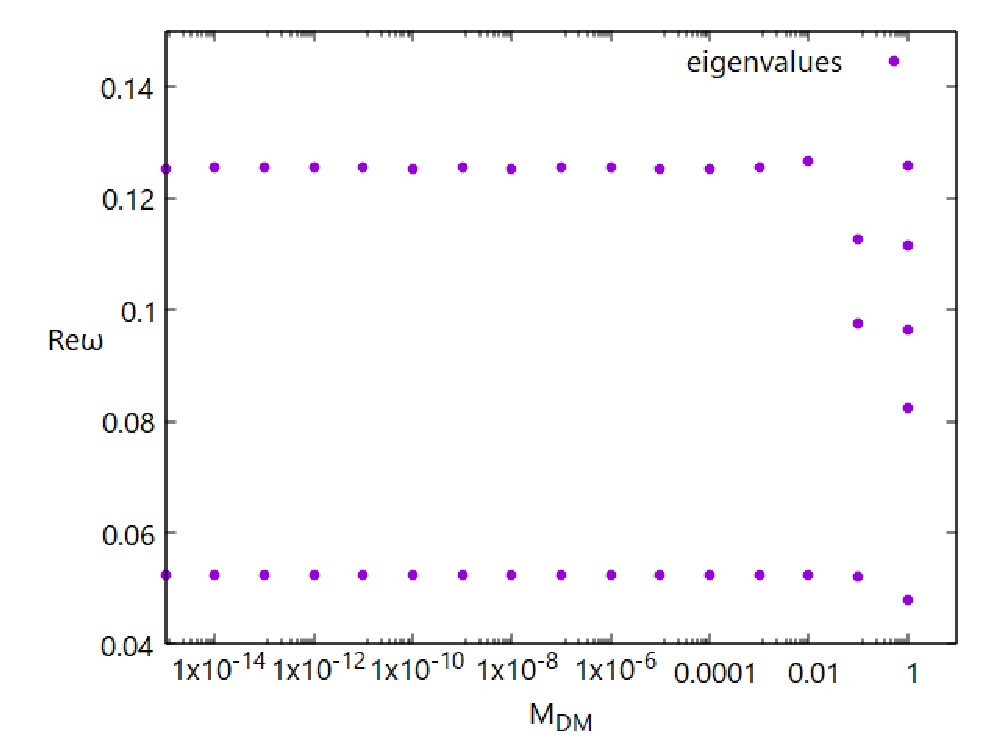}
\caption{Real part of eigenvalue $\omega$ as a function of $M_{_\mathrm{DM}}$(SFDM,$\ell=1,m=1,a=0.5,\alpha=1$).}
\label{sfdmMDMRe}
\end{minipage}
\begin{minipage}{1.0\columnwidth}
\centering
\includegraphics[scale=0.5]{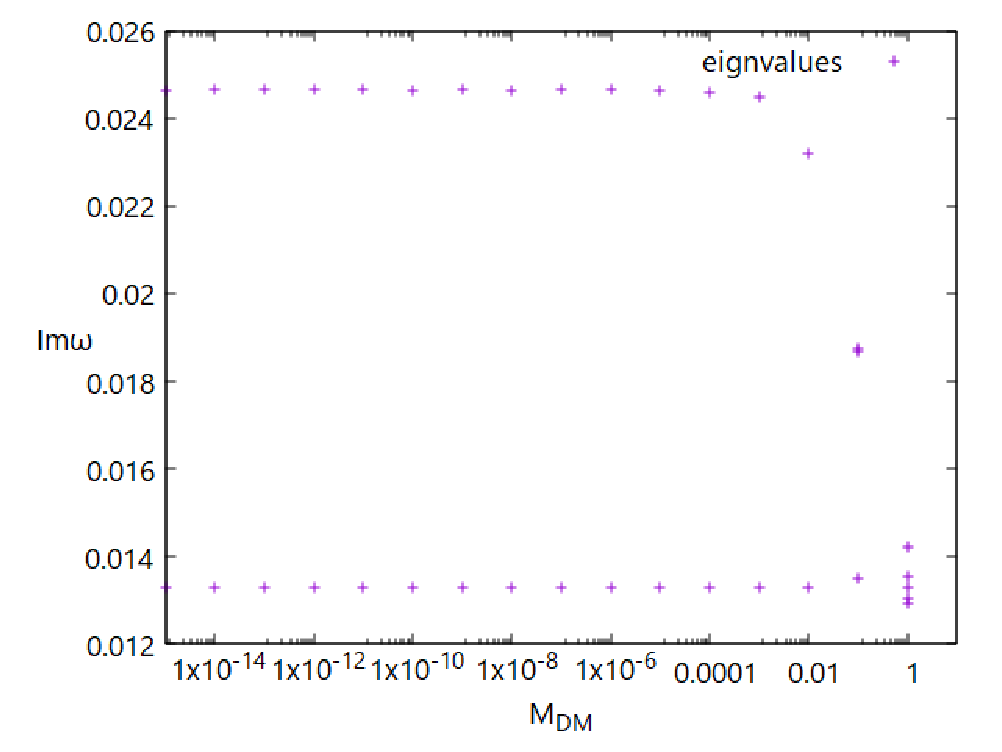}
\caption{Imginary part of eigenvalue $\omega$ as a function of $M_{_\mathrm{DM}}$(SFDM,$\ell=1,m=1,a=0.5,\alpha=1$).}
\label{sfdmMDMIm}
\end{minipage}
\end{figure*}

Fig.\ref{sfdmMDMRe} and \ref{sfdmMDMIm} show Re[$\omega$] and Im[$\omega$] for each $M_{_\mathrm{DM}}$ in the case of SFDM. For SFDM, the number of modes and amplitudes also remain almost unchanged for small $M_{_\mathrm{DM}}$, but the number of modes for large $M_{_\mathrm{DM}}$ differs from CDMs. Furthermore, the values for large $M_{_\mathrm{DM}}$ differ from those of CDM, but the values for small $M_{_\mathrm{DM}}$ are close to those of CDM. In other words, the difference between CDM and SFDM appears only when MDM is somewhat large.

\begin{figure*}
\centering
\begin{minipage}{1.0\columnwidth}
\centering
\includegraphics[scale=0.5]{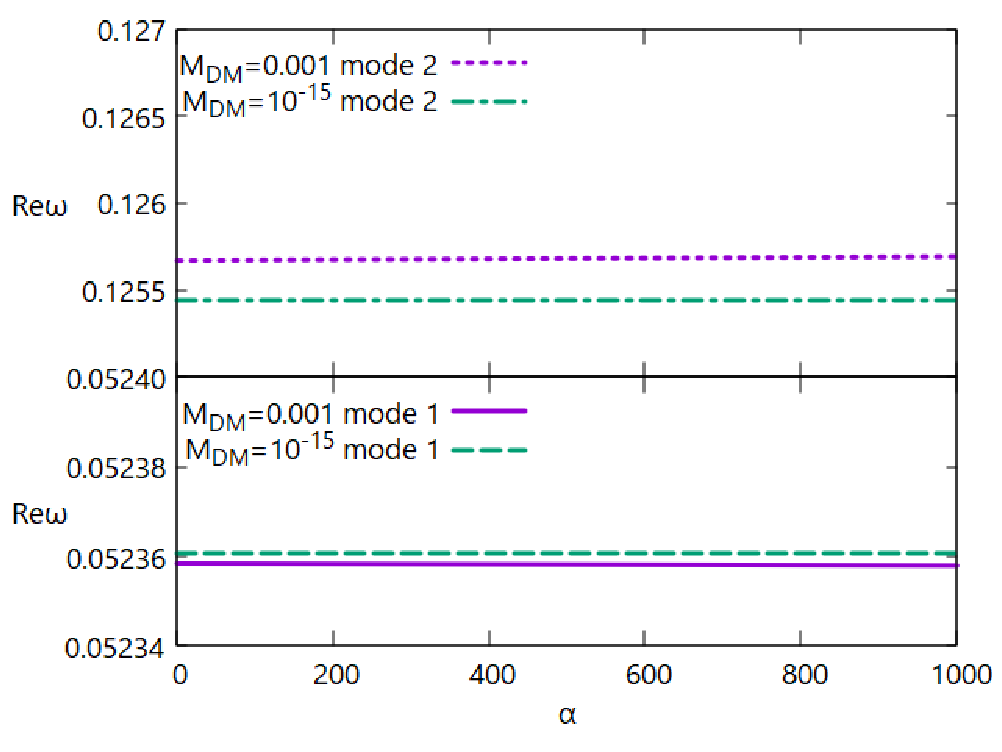}
\caption{Real part of eigenvalue $\omega$ as a function of $\alpha$ for $M_{_\mathrm{DM}}=10^{-15}, 0.001$ (SFDM,$\ell=1,m=1,a=0.5,R=100$)}
\label{sfdmalphaRe}
\end{minipage}
\begin{minipage}{1.0\columnwidth}
\centering
\includegraphics[scale=0.5]{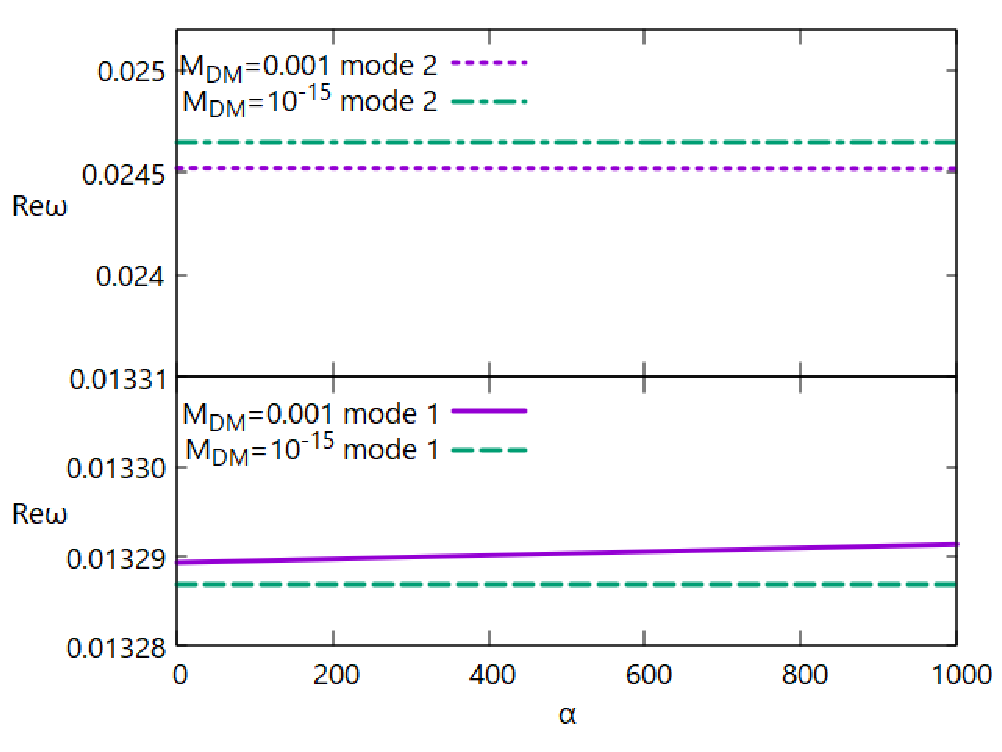}
\caption{Imginary part of eigenvalue $\omega$ as a function of $\alpha$ for $M_{_\mathrm{DM}}=10^{-15}, 0.001$ (SFDM,$\ell=1,m=1,a=0.5,R=100,$)}
\label{sfdmalphaIm}
\end{minipage}
\end{figure*}

\begin{figure*}
\centering
\begin{minipage}{1.0\columnwidth}
\centering
\includegraphics[scale=0.5]{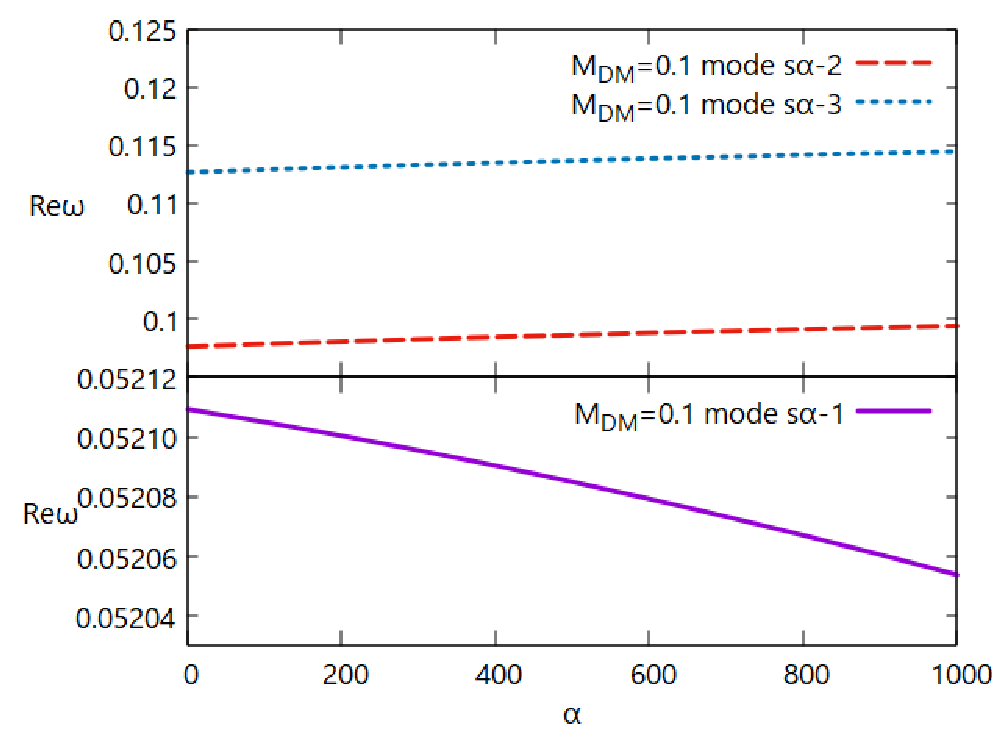}
\caption{Real part of eigenvalue $\omega$ as a function of $\alpha$(SFDM,$\ell=1,m=1,a=0.5,R=100,M_{_\mathrm{DM}}=0.1$)}
\label{sfdmalphaMDM0.1Re}
\end{minipage}
\begin{minipage}{1.0\columnwidth}
\centering
\includegraphics[scale=0.5]{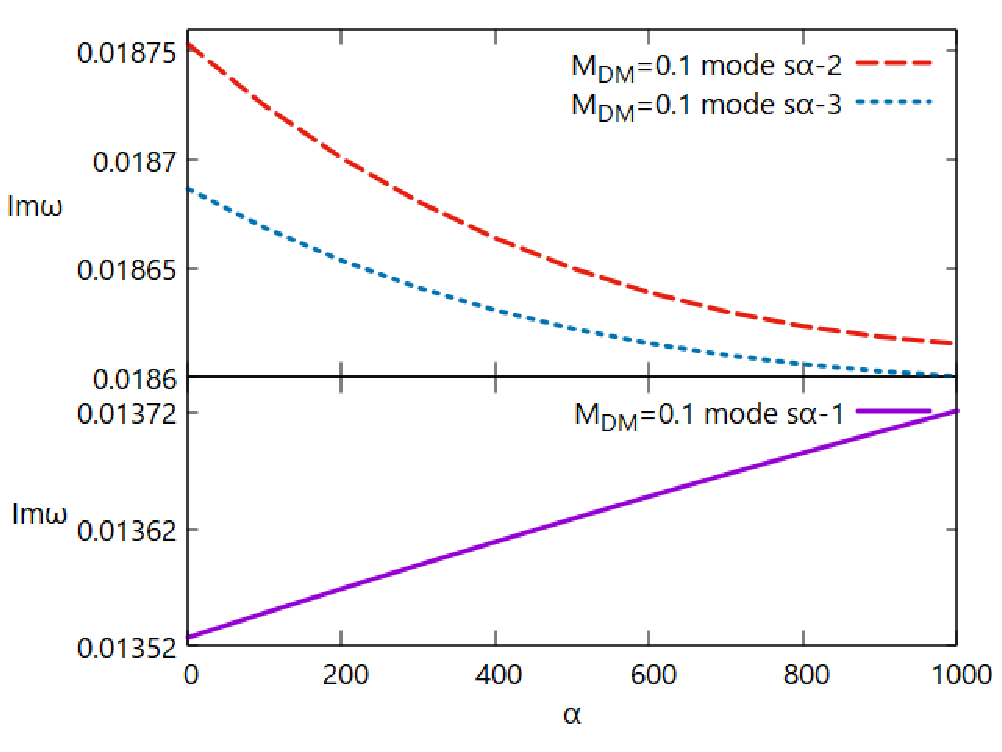}
\caption{Imginary part of eigenvalue $\omega$ as a function of $\alpha$(SFDM,$\ell=1,m=1,a=0.5,R=100,M_{_\mathrm{DM}}=0.1$)}
\label{sfdmalphaMDM0.1Im}
\end{minipage}
\end{figure*}

Fig.\ref{sfdmalphaRe}$-$\ref{sfdmalphaMDM0.1Im} show Re[$\omega$] and Im[$\omega$] for each $\alpha$ with $M_{_\mathrm{DM}}=10^{-15} ,0.001, 0.1$. In the case of SFDM, the amplitude change due to $\alpha$ is smaller than that of CDM. Also, unlike CDM, the number of modes does not increase unless $M_{_\mathrm{DM}} \geq 0.1$.

\begin{figure*}
\centering
\begin{minipage}{1.0\columnwidth}
\centering
\includegraphics[scale=0.5]{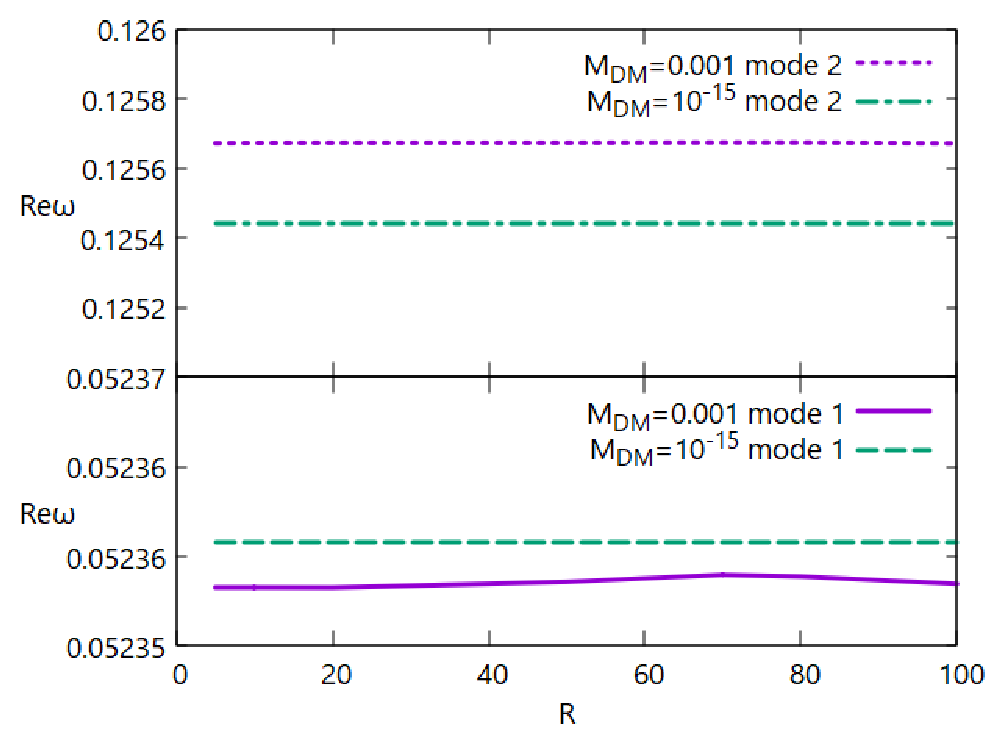}
\caption{Real part of eigenvalue $\omega$ when $R$ is varied up to $100$ for $M_{_\mathrm{DM}}=10^{-15}, 0.001$ (SFDM,$\alpha=1,\ell=1,m=1,a=0.5$).}
\label{sfdmRRe}
\end{minipage}
\begin{minipage}{1.0\columnwidth}
\centering
\includegraphics[scale=0.5]{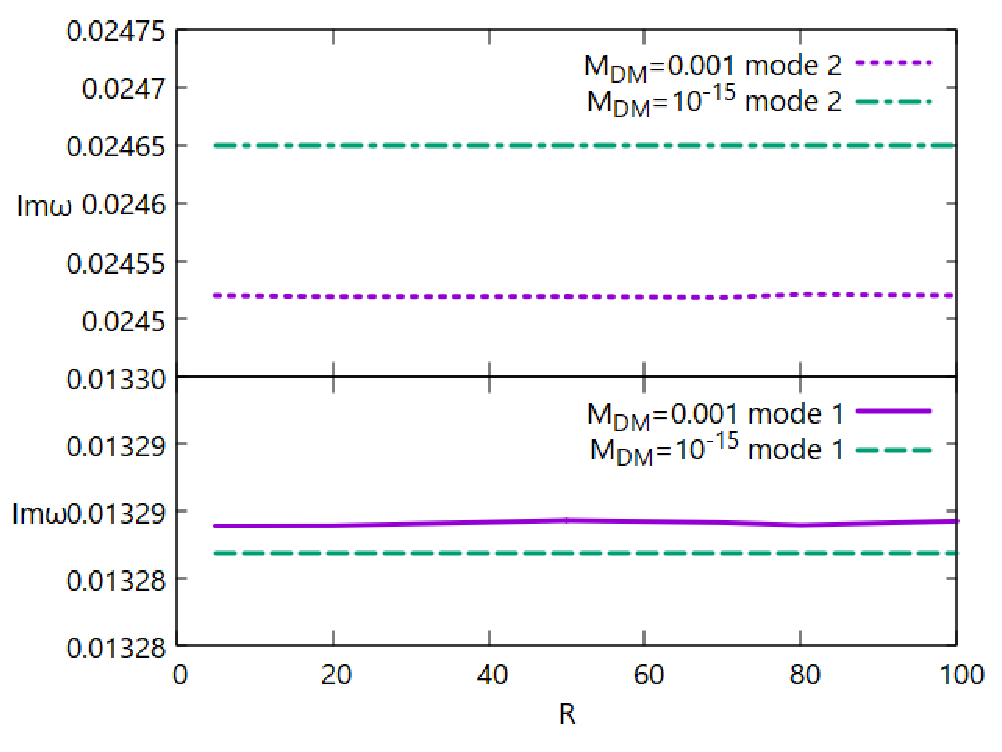}
\caption{Imginary part of eigenvalue $\omega$ when $R$ is varied up to $100$ for $M_{_\mathrm{DM}}=10^{-15}, 0.001$ (SFDM,$\alpha=1,\ell=1,m=1,a=0.5$).}
\label{sfdmRIm}
\end{minipage}
\end{figure*}

\begin{figure*}
\centering
\begin{minipage}{1.0\columnwidth}
\centering
\includegraphics[scale=0.5]{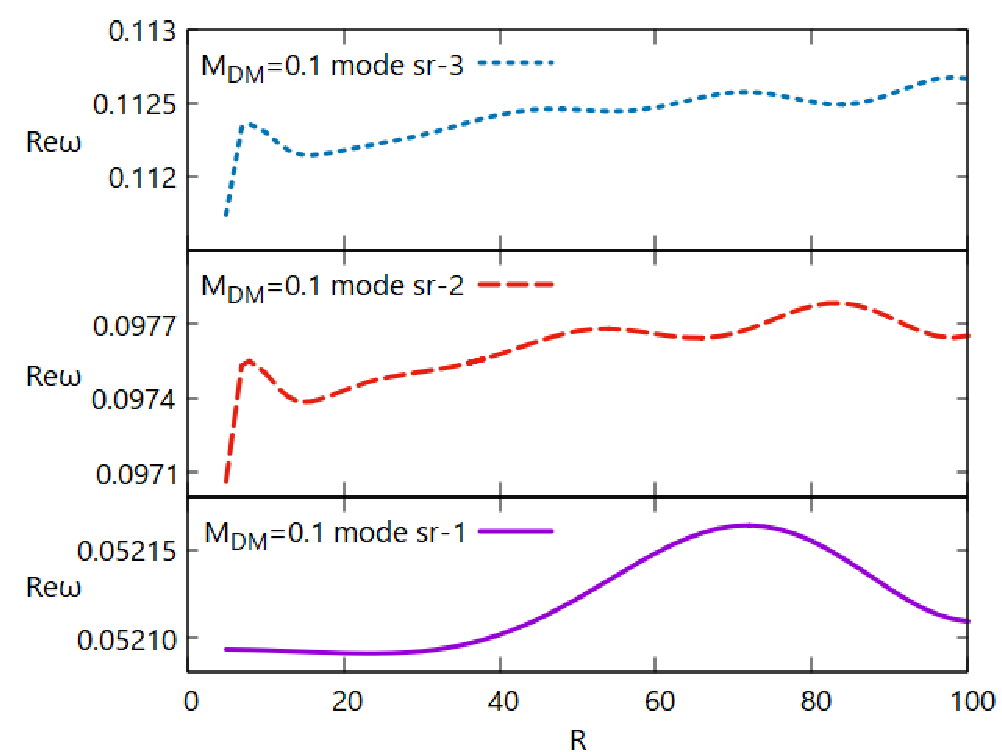}
\caption{Real part of eigenvalue $\omega$ when $R$ is varied up to $100$ at $M_{_\mathrm{DM}}=0.1$ (SFDM,$\alpha=1,\ell=1,m=1,a=0.5$).}
\label{sfdmRMDM0.1Re}
\end{minipage}
\begin{minipage}{1.0\columnwidth}
\centering
\includegraphics[scale=0.5]{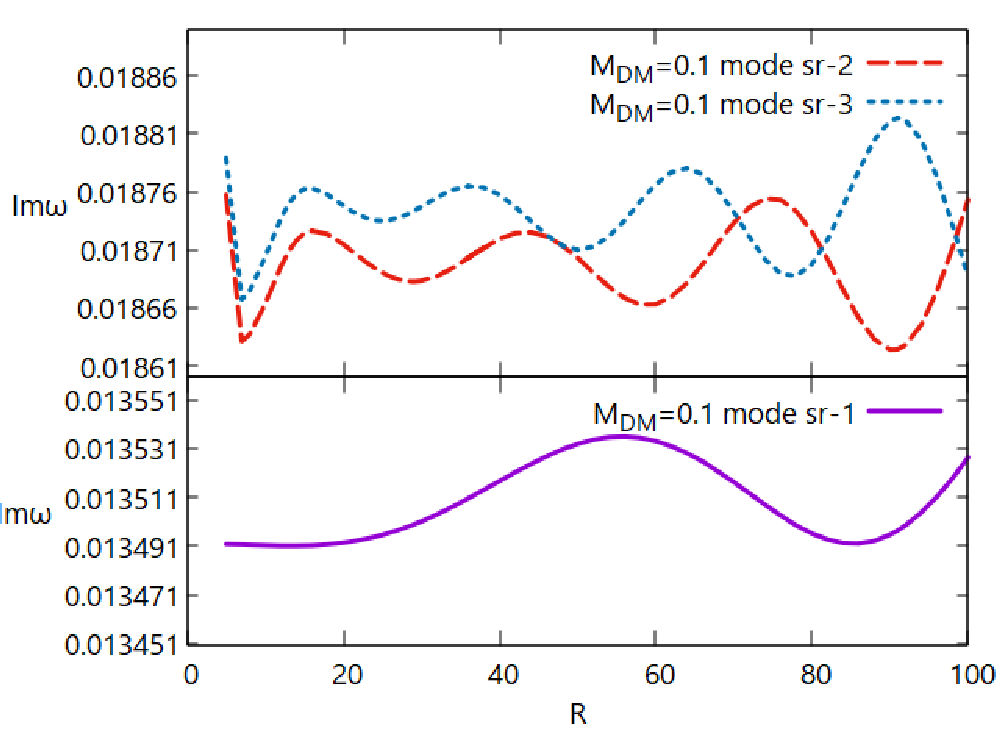}
\caption{Imginary part of eigenvalue $\omega$ when $R$ is varied up to $100$ at $M_{_\mathrm{DM}}=0.1$(SFDM,$\alpha=1,\ell=1,m=1,a=0.5$).}
\label{sfdmRMDM0.1Im}
\end{minipage}
\end{figure*}

Moreover, changes in $R$ cause  significant changes in amplitude. Fig.\ref{sfdmRRe} $-$\ref{sfdmRMDM0.1Im} show the changes of Re[$\omega$] and Im[$\omega$] when $R$ is varied for $M_{_\mathrm{DM}}=10^{-15}, 0.001, 0.1$. While the CDM case shows a monotonic increase or decrease, the SFDM case is different: in large $M_{_\mathrm{DM}}$, both Re[$\omega$] and Im[$\omega$] show periodic changes when R is varied. This may be due to the inclusion of trigonometric functions in the effective mass and thus in the metric.
\par

\section{Discussion and Conclusion}
\label{Consec}
One of the objectives of this study is to investigate whether scalarization occurs in BH surrounded by DM halo in scalar tensor theory. Due to the effective mass created by the coupling of the scalar and matter fields, perturbations that do not occur in GR can occur. In both the CDM and SFDM cases, scalarization occurs only when $\alpha$ is negative to some extent, and no scalarization occurs when $\alpha >0$. This is the same situation as that encountered in the normal matter DEF model, where scalarization is severely limited by the binary-pulsar limit\cite{2024RvMP...96a5004D}. We also checked whether scalarization occurs not only for the small halo model but also for specific galaxies, and the results are the same. However, under very weak constraints \cite{1996PhRvD..54.1474D}, it may be possible for small DM halos to be able to have scalarization.
\par
In order to know more general properties of DM halo scalarization, we also  invegestigate the dependence of $R_s(R)$, $M_{_\mathrm{DM}}$($\rho_c$) and $M$. The results show that there are dependences not only on $M_{_\mathrm{DM}}$ but also on the central BH mass. The degree of scalrization (absolute value of $S$) is larger for larger $M_{_\mathrm{DM}}$ and smaller central BH mass $M$. In combination with the DM halo size issue, scalarization may be more likely to occur in smaller systems rather than in galactic halo sized BH-DM systems.
\par
Our result is that scalarization occurs in BHs with DM halos, but we have not studied the consequence: as a result of scalarization, the system consisting of a DM halo and a BH becomes a new system with scalar hairs, but this process is beyond the scope of this paper. The transition to a new state due to scalarization may also change the structure of the halo.\par
Our other purpose is to investigate the superradiant instability that occurs when the sign of the effective mass of the scalar field changes. As a result, we confirmed that the superradiant instability occurs in both the CDM and SFDM cases for a certain rotation parameter $a$, even in the case of a slow-rotating BH. In this case, it is confirmed that the larger $a$ has the larger Im[$\omega$]. Also, the dependence of unstable eigenmode on $\alpha$ is found to be significantly different between CDM and SFDM.
\par
However, the most important influence on the superradiant instability is the structure of the halo, especially the mass of the DM,  $M_{_\mathrm{DM}}$.
The number of unstable modes are affected by $M_{_\mathrm{DM}}$. The larger $M_{_\mathrm{DM}}$ is, the more  unstable modes exist, but the absolute value of Im[$\omega$] becomes smaller. This phenomenon is an intrinsic property of the BH-DM halo system. Heavier halos than some extent have the effect of reducing instability. The same can be read from the dependence on the size of the halo (however, in SFDM, instability  fluctuates periodically as $R$ increases, so it cannot be described in the same way as in CDM.).
\par
We have considered models with small size halos but the actual galactic halo is much larger than them. We could read various properties of the system , but our model is a kind of toy model. However, when considering the matter-coupling superradiance in scalar-tensor theory, the boundary condition at infinity is proportional to $e^{ikx}$. In this case, if the numerical infinity is too large, the boundary condition diverges. To prevent this we have restricted the size of the halo, but this problem seriously limits the calculations of the galctic halo of the actual size. Unless this problem is solved, it will be difficult to compute the superradiance of the actual halo in scalar tensor theory.\par

\clearpage
\appendix

\section{\label{effmass}effective mass}
The spacetime metric of BH-DM system is
\begin{align}
ds^2=&-\left(1-\frac{r^2+2Mr-r^2f(r)}{\Sigma^2}\right)dt^2\notag \\
&+\frac{\Sigma^2}{\varDelta}dr^2+\Sigma^2 d\theta^2+\frac{A\sin^2\theta}{\Sigma^2}d\varphi^2 \notag \\
&-\frac{2(r^2+2Mr-r^2f(r))a\sin^2\theta}{\Sigma^2}d\varphi dt ,
\end{align}
where
\begin{align}
A&=(r^2+a^2)^2-a^2\varDelta \sin^2 \theta , \\
B&=r^2+2Mr-r^2f(r) ,\\
\Sigma^2&=r^2+a^2cos\theta , \\
\varDelta&=r^2 f(r)-2Mr+a^2 \approx (r-r_{+})(r-r{-}) , 
\end{align}

\begin{widetext}
\begin{align}
g_{\mu \nu}  
=\begin{pmatrix}
                   -(1-\frac{r^2+2Mr-r^2f(r)}{\Sigma^2}) &0&0&-\frac{ar(r^2+2Mr-r^2f(r))}{\varDelta\Sigma^2} \\
                   0 &\frac{\Sigma^2}{\varDelta} &0 &0 \\
                   0 &0&\Sigma^2&0 \\
                   -\frac{ar(r^2+2Mr-r^2f(r))}{\varDelta\Sigma^2}&0&0& \frac{A \sin^2 \theta}{\Sigma^2} 
                    \end{pmatrix} ,
\end{align}
\end{widetext}
\begin{widetext}
\begin{align}
g^{\mu \nu}
=\begin{pmatrix}
                   -\frac{A}{\varDelta \Sigma^2} &0&0&-\frac{a\sin^2 \theta (r^2+2Mr-r^2f(r))}{\Sigma^2} \\
                   0 &\frac{\varDelta}{\Sigma^2} &0 &0 \\
                   0 &0&\frac{1}{\Sigma^2}&0 \\
                   -\frac{a\sin^2 \theta (r^2+2Mr-r^2f(r))}{\Sigma^2}&0&0& \frac{-(r^2+2Mr-r^2f(r))+\Sigma^2}{\varDelta \Sigma^2 \sin^2\theta} 
                    \end{pmatrix} .
\end{align}
\end{widetext}

From \cite{misner2017gravitation},Ricci tenor is generally represented by 
\begin{equation}
R_{\alpha \beta} =-(\log(\sqrt{|g|}))_{,\alpha \beta}+\varGamma^\gamma_{\alpha \beta,\gamma}+\log(\sqrt{|g|})_{,\gamma}\varGamma^\gamma_{\alpha \beta,\gamma}-\varGamma^\gamma_{\beta \delta}\varGamma^\delta_{\alpha \gamma},
\end{equation}
where
\begin{align}
g&=\det g_{\mu \nu}=-\Sigma^4 \sin^2\theta , \\
\sqrt{|g|}&=\Sigma^2 \sin\theta , \\
(\log(\sqrt{|g|}))_{,r}&=\frac{\Sigma^2_{,r}}{\Sigma^2} ,\\
(\log(\sqrt{|g|}))_{,\theta}&=\frac{\Sigma^2_{,\theta}}{\Sigma^2}+\frac{\cos \theta}{\sin \theta} ,\\
(\log(\sqrt{|g|}))_{,rr}&=\frac{\Sigma^2_{,rr}}{\Sigma^2}-\frac{(\Sigma^2_{,r})^2}{\Sigma^4} ,\\
(\log(\sqrt{|g|}))_{,\theta \theta}&=\frac{\Sigma^2_{,\theta \theta}}{\Sigma^2}-\frac{(\Sigma^2_{,\theta})^2}{\Sigma^4}-\frac{1}{\sin^2\theta} .
\end{align}

By the slow rotation approximation $a^2 \rightarrow 0$, these elements become
\begin{equation}
\varDelta \simeq r^2f(r)-2Mr ,\quad \Sigma^2  \simeq r^2 ,\quad A\simeq r^4 ,\notag
\end{equation}
\begin{equation}
B\simeq r^2-\varDelta ,\quad \Sigma^2_{,r} \simeq 2r ,\quad \Sigma^2_{,rr} \simeq 2 , \notag
\end{equation}
\begin{equation}
\Sigma^2_{,\theta} \simeq0 ,\quad \Sigma^2_{,\theta\theta } \simeq0 ,\quad A_{,r} \simeq 4r^3 , \notag
\end{equation}
\begin{equation}
A_{,\theta} \simeq 0 , \quad A_{,rr} \simeq 12r^2 , \quad A_{,\theta \theta} \simeq 0 , \notag
\end{equation}
\begin{equation}
B_{,r}\simeq 2r-\varDelta_{,r} ,\quad B_{,rr}\simeq2-\varDelta_{,rr} , \notag
\end{equation}
\begin{equation}
\varDelta_{,r}=2rf(r)+r^2f'(r)-2M , \notag
\end{equation}
\begin{equation}
\varDelta_{,rr}=2f(r)+4rf'(r)+r^2f''(r) ,\notag
\end{equation}

By using these elements, we can caluculate non-zero componets of $\varGamma^\gamma_{\alpha \beta}$:
\begin{align}
\varGamma^t_{t r}&=\frac{1}{2}g^{tt}g_{tt,r}=-\frac{A(B_{,r}\Sigma^2-B\Sigma^2_r)}{2 \varDelta \Sigma^6} \rightarrow \frac{\varDelta_{,r}}{2\varDelta}-\frac{1}{r},\\
\varGamma^t_{t \theta}&=\frac{1}{2}g^{tt}g_{tt,r}=\frac{AB\Sigma^2_{,\theta}}{2\varDelta \Sigma^6} \rightarrow0 , \\
\varGamma^r_{t t}&=-\frac{1}{2}g^{rr}g_{tt,r}=-\frac{\varDelta(B_{,r}\Sigma^2-B\Sigma_{,r})}{2\Sigma^6}\rightarrow \frac{\varDelta \varDelta_{,r}}{2r^4}-\frac{\varDelta^2}{r^5} ,\\
\varGamma^r_{r r}&=\frac{1}{2}g^{rr}g_{rr,r}=\frac{1}{2}(\frac{\Sigma^2_{,r}}{\Sigma^2}-\frac{\varDelta_{,r}}{\varDelta})\rightarrow \frac{1}{r}-\frac{\varDelta_{,r}}{2\varDelta} , \\
\varGamma^r_{t \varphi}&=-\frac{1}{2}g^{rr}g_{t \varphi,r}=\frac{\varDelta(B_{,r}\Sigma^2-B\Sigma^2_{,r})}{2\Sigma^6}a \sin^2\theta \notag \\
&\rightarrow \frac{\varDelta^2}{r^5}a\sin^2\theta -\frac{\varDelta \varDelta_{,r}}{2r^4}a\sin^2\theta , 
\end{align}
\begin{align}
\varGamma^r_{\theta r}&=\frac{1}{2}g^{rr}g_{rr,\theta}=\frac{\Sigma^2_{,\theta}}{2\Sigma^2} \rightarrow0 , \\
\varGamma^r_{\theta \theta}&=-\frac{\varDelta \Sigma^2_{,r}}{2\Sigma^2}\rightarrow -\frac{\varDelta}{r} ,\\
\varGamma^r_{\varphi \varphi}&=-\frac{\varDelta(A_{,r}\Sigma^2-A\Sigma^2_{,r})}{2\Sigma^6} \sin^2 \theta \rightarrow -\frac{\varDelta}{r} , \\
\varGamma^\theta_{t t}&=-\frac{1}{2}g^{\theta \theta}g_{tt,\theta}=\frac{B\Sigma^2_{,\theta}}{2\Sigma^6}\rightarrow0 ,\\
\varGamma^\theta_{t \varphi}&=-\frac{1}{2}g^{\theta \theta}g_{t \varphi,\theta}=\frac{aB(2\sin\theta \cos\theta \Sigma^2-\sin^2 \Sigma^2_{,\theta})}{2\Sigma^6}\notag\\
&\rightarrow \frac{aB\sin\theta \cos\theta}{r^4} , \\
\varGamma^\theta_{r r}&=-\frac{1}{2}g^{\theta \theta}g_{rr,\theta}=-\frac{\Sigma^2_{,\theta}}{2\varDelta \Sigma^2}\rightarrow0 ,\\
\varGamma^\theta_{r \theta}&=\frac{1}{2}g^{\theta \theta}g_{\theta \theta ,r}=\frac{\Sigma^2_{,r}}{2\Sigma^2}\rightarrow \frac{1}{r} ,\\
\varGamma^\theta_{\theta \theta}&=\frac{1}{2}g^{\theta \theta}g_{\theta \theta,\theta}=\frac{\Sigma^2_{,\theta}}{2\Sigma^2}\rightarrow0 , \\
\varGamma^\theta_{\varphi \varphi}&=-\frac{1}{2}g^{\theta \theta}g_{\varphi \varphi,\theta}\notag\\
&=-\frac{(A_{,\theta}\sin^2\theta+2A\sin\theta \cos\theta)\Sigma^2-A\sin^2\theta \Sigma^2_{,\theta}}{2\Sigma^6}\notag\\
&\rightarrow-\sin\theta\cos\theta , \\
\varGamma^\varphi_{t r}&=\frac{1}{2}g^{\varphi \varphi}g_{\varphi t,r}+\frac{1}{2}g^{\varphi t}g_{tt,r}\notag\\
&=\frac{a(B \Sigma^2_{,r}-B_{,r}\Sigma^2)}{2\varDelta \Sigma^4}\rightarrow \frac{a\varDelta_{,r}}{2\varDelta r^2}-\frac{a}{r^3} ,\\
\varGamma^\varphi_{t \theta}&=\frac{1}{2}g^{\varphi t}g_{t t,\theta}+\frac{1}{2}g^{\varphi \varphi}g_{\varphi t,\theta}\notag\\
&=\frac{a B^2 \Sigma^2_{,\theta}}{2\varDelta \Sigma^6}-\frac{aB(-B+\Sigma^2)(2\sin\theta\cos\theta \Sigma^2-\sin^2\theta \Sigma_{,\theta})}{2\varDelta \Sigma^6 \sin^2\theta}\notag\\
&\rightarrow \frac{a\varDelta \cos \theta}{r^4\sin \theta}-\frac{a\cos \theta}{r^2\sin \theta} ,\\
\varGamma^\varphi_{r \varphi}&=\frac{1}{2}g^{\varphi \varphi}g_{\varphi \varphi,r}=\frac{(-B+\Sigma^2)(A_{,r}\Sigma^2-A \Sigma^2_{,r})}{2\varDelta \Sigma^6}\rightarrow \frac{1}{r} ,\\
\varGamma^\varphi_{\theta \varphi}&=\frac{1}{2}g^{\varphi \varphi}g_{\varphi \varphi,\theta}\notag\\
&=\frac{(-B+\Sigma^2)((A_{,\theta}\sin^2\theta+2A\sin\theta \cos\theta)\Sigma^2-\sin^2\theta \Sigma^2_{,\theta})}{2\varDelta \Sigma^6\sin^2\theta}\notag\\
&\rightarrow\frac{\cos\theta}{\sin\theta} .
\end{align}

The derivatives of $\varGamma^\gamma_{\alpha \beta}$ are
\begin{align}
\varGamma^r_{t t,r}&=-\frac{\varDelta_{,r} B_{,r}}{2\Sigma^4} -\frac{\varDelta B_{,rr}}{2\Sigma^4} +\frac{\varDelta B_{,r}\Sigma^2_{,r}}{\Sigma^6}\notag+\frac{\varDelta_{,r} B \Sigma^2_{,r}}{2\Sigma^6}\notag\\
&+\frac{\varDelta B_{,r} \Sigma^2_{,r}}{2\Sigma^6}+\frac{\varDelta B \Sigma^2_{,rr}}{2\Sigma^6}+\frac{3\varDelta B (\Sigma^2_{,r})^2}{2\Sigma^8} ,\\
\varGamma^\theta_{t t,\theta}&=\frac{B\Sigma^2_{,\theta \theta}}{2\Sigma^6}\rightarrow 0 ,
\end{align}
\begin{align}
\varGamma^r_{r r,r}&=\frac{1}{2}\left(\frac{\Sigma^2_{,rr}}{\Sigma^2}-\frac{(\Sigma^2_{,r})^2}{\Sigma^4} -\frac{\varDelta_{,rr}}{\varDelta}+\frac{(\varDelta_{,r})^2}{\varDelta^2} \right)\notag\\
&\rightarrow \frac{1}{2}(-\frac{2}{r^2}-\frac{\varDelta_{,rr}}{\varDelta}+\frac{(\varDelta_{,r})^2}{\varDelta^2}) ,\\
\varGamma^\theta_{r r,\theta}&=-\frac{1}{2}\left(\frac{\Sigma^2_{,\theta \theta}}{\varDelta \Sigma^2}-\frac{(\Sigma_{,\theta})^2}{\varDelta \Sigma^4}\right)\rightarrow 0 ,\\
\varGamma^r_{\theta \theta,r}&=-\frac{\varDelta_{,r} \Sigma^2_{,r}}{2\Sigma^2}-\frac{\varDelta \Sigma^2_{,rr}}{2\Sigma^2}+\frac{\varDelta (\Sigma_{,r})^2}{2\Sigma^4}\rightarrow -\frac{\varDelta_{,r}}{r}+\frac{\varDelta}{r^2} , \\
\varGamma^\theta_{\theta \theta,\theta}&=\frac{\Sigma^2_{,\theta \theta}}{2\Sigma^2}-\frac{(\Sigma^2_{,\theta })^2}{2\Sigma^4}\rightarrow 0,\\
\varGamma^r_{\varphi \varphi,r}&=\Big(-\frac{\varDelta_{,r}A_{,r}}{2\Sigma^4} -\frac{\varDelta A_{,rr}}{2\Sigma^4} +\frac{\varDelta A_{,r} \Sigma^2_{,r}}{\Sigma^6}\notag\\
&+\frac{\varDelta_{,r} A \Sigma^2_{,r}}{2\Sigma^6}+\frac{\varDelta A_{,r} \Sigma^2_{,r}}{2\Sigma^6}+ \frac{\varDelta A \Sigma^2_{,rr}}{2\Sigma^6}\notag\\
&- 3\frac{\varDelta A (\Sigma^2_{,r})^2}{2\Sigma^6}\Big)\sin^2 \theta\notag\\ &\rightarrow \frac{\varDelta}{r^2}\sin^2\theta ,\\
\varGamma^\theta_{\varphi \varphi,\theta}&=-\frac{A_{,\theta \theta}\sin^2 \theta +2A_{,\theta}\sin \theta \cos \theta+2A_{,\theta}\sin \theta \cos \theta}{2\Sigma^4}\notag\\
&-\frac{2A\cos^2 \theta-2A\sin^2 \theta}{2\Sigma^4}\notag\\
&+\frac{(A_{,\theta}\sin^2 \theta+2A\sin \theta \cos \theta)\Sigma^2_{,\theta}}{\Sigma^6}\notag\\
&+\frac{A_{,\theta}\sin^2 \theta \Sigma^2_{,\theta}+2A\sin \theta \cos \theta \Sigma^2_{,\theta}+A\sin^2 \theta \Sigma^2_{\theta \theta} }{2\Sigma^6}\notag\\
&-\frac{3A\sin^2 \theta (\Sigma^2_{,\theta})^2}{2\Sigma^8}\notag\\
&\rightarrow -\cos^2\theta+\sin^2\theta .
\end{align}
Then the non-zero components of Ricci tensor are,
\begin{align}
R_{tt}=&\varGamma^r_{t t,r}+\varGamma^\theta_{t t,\theta}+(\log(\sqrt{|g|}))_{,r}\varGamma^r_{t t}+(\log(\sqrt{|g|}))_{,\theta}\varGamma^\theta_{t t} \notag\\
&-2\varGamma^{t}_{tr}\varGamma^{r}_{tt}-2\varGamma^{t}_{t\theta}\varGamma^{\theta}_{tt}-2\varGamma^{\theta}_{t\varphi}\varGamma^{\varphi}_{t\theta}-2\varGamma^{r}_{t\varphi}\varGamma^{\varphi}_{tr}\notag\\
&\rightarrow \varGamma^r_{t t,r}+(\log(\sqrt{|g|}))_{,r}\varGamma^r_{t t}-2\varGamma^{t}_{tr} ,\\
R_{rr}=&-(\log(\sqrt{|g|}))_{,rr}+\varGamma^r_{rr,r}+\varGamma^\theta_{rr,\theta}\notag\\
&+(\log(\sqrt{|g|}))_{,r}\varGamma^r_{rr}+(\log(\sqrt{|g|}))_{,\theta}\varGamma^\theta_{rr}\notag\\
&-\varGamma^t_{rt}\varGamma^t_{rt}-\varGamma^r_{rr}\varGamma^r_{rr}-2\varGamma^r_{r\theta}\varGamma^\theta_{rr}-\varGamma^\theta_{r\theta}\varGamma^\theta_{r\theta}-\varGamma^\varphi_{r\varphi}\varGamma^\varphi_{r\varphi}\notag\\
&\rightarrow -(\log(\sqrt{|g|}))_{,rr}+\varGamma^r_{rr,r}+(\log(\sqrt{|g|}))_{,r}\varGamma^r_{rr}\notag\\
&-\varGamma^t_{rt}\varGamma^t_{rt}-\varGamma^r_{rr}\varGamma^r_{rr}-\varGamma^\theta_{r\theta}\varGamma^\theta_{r\theta}-\varGamma^\varphi_{r\varphi}\varGamma^\varphi_{r\varphi} ,\\
R_{\theta \theta}=&-(\log(\sqrt{|g|}))_{,\theta \theta}+\varGamma^r_{\theta \theta,r}+\varGamma^\theta_{\theta \theta,\theta}\notag\\
&+(\log(\sqrt{|g|}))_{,r}\varGamma^r_{\theta \theta}+(\log(\sqrt{|g|}))_{,\theta}\varGamma^\theta_{\theta \theta}\notag\\
&-\varGamma^t_{\theta t}\varGamma^t_{\theta t}-\varGamma^r_{\theta r}\varGamma^r_{\theta r}-2\varGamma^r_{\theta \theta}\varGamma^\theta_{r\theta}-\varGamma^\theta_{\theta \theta}\varGamma^\theta_{\theta \theta}-\varGamma^\varphi_{\theta \varphi}\varGamma^\varphi_{\theta \varphi}\notag\\
&\rightarrow -(\log(\sqrt{|g|}))_{,\theta \theta}+\varGamma^r_{\theta \theta,r}+(\log(\sqrt{|g|}))_{,r}\varGamma^r_{\theta \theta}\notag\\
&-2\varGamma^r_{\theta \theta}\varGamma^\theta_{r\theta}-\varGamma^\varphi_{\theta \varphi}\varGamma^\varphi_{\theta \varphi} ,
\end{align}
\begin{align}
R_{\varphi \varphi}=&\varGamma^r_{\varphi \varphi,r}+\varGamma^\theta_{\varphi \varphi,\theta}+(\log(\sqrt{|g|}))_{,r}\varGamma^r_{\varphi \varphi}\notag\\
&+(\log(\sqrt{|g|}))_{,\theta}\varGamma^\theta_{\varphi \varphi}\notag\\
&-2\varGamma^r_{\varphi \varphi}\varGamma^\varphi_{\varphi r}-2\varGamma^\theta_{\varphi \varphi}\varGamma^\varphi_{\varphi \theta}\notag\\
&\rightarrow \varGamma^r_{\varphi \varphi,r}+\varGamma^\theta_{\varphi \varphi,\theta}+(\log(\sqrt{|g|}))_{,r}\varGamma^r_{\varphi \varphi}\notag\\
&+(\log(\sqrt{|g|}))_{,\theta}\varGamma^\theta_{\varphi \varphi}-2\varGamma^r_{\varphi \varphi}\varGamma^\varphi_{\varphi r}-2\varGamma^\theta_{\varphi \varphi}\varGamma^\varphi_{\varphi \theta} ,\\
R_{t \varphi}=&\varGamma^r_{t \varphi,r}+\varGamma^\theta_{t \varphi,\theta}+(\log(\sqrt{|g|}))_{,r}\varGamma^r_{t \varphi}+(\log(\sqrt{|g|}))_{,\theta}\varGamma^\theta_{t\varphi}\notag\\
&-\varGamma^r_{t \varphi}\varGamma^t_{tr}-\varGamma^\theta_{\varphi \varphi}\varGamma^\varphi_{t \theta}-\varGamma^\varphi_{\varphi r}\varGamma^r_{t \varphi}-\varGamma^\varphi_{\varphi \theta}\varGamma^\theta_{t \varphi} \notag\\
&\propto a .
\end{align}
Ricci scalr is,
\begin{equation}
R=g^{tt}R_{tt}+g^{rr}R_{rr}+g^{\theta \theta}R_{\theta \theta}+g^{\varphi \varphi}R_{\varphi \varphi}+2g^{t \varphi}R_{t \varphi} ,
\end{equation}
However, because of  $g^{t \varphi} \propto a$, $R_{t \varphi} \propto a$, $g^{t \varphi}R_{t \varphi}\propto a^2 \rightarrow 0$.
Therefore, in the slow rotation limit,
\begin{align}
R\simeq& -\frac{r^2}{\varDelta }\Big[\frac{(\varDelta_{,r})^2}{2r^4}+\frac{\varDelta \varDelta_{,rr}}{2r^4}-\frac{4\varDelta \varDelta_{,r}}{r^5}+\frac{5 \varDelta^2}{r^6}\notag\\
&+\frac{2}{r}\Big(\frac{\varDelta \varDelta_{,r}}{2r^4}-\frac{\varDelta^2}{r^5}\Big)-2\Big(\frac{\varDelta \varDelta_{,r}}{2r^4}-\frac{\varDelta^2}{r^5}\Big)\Big(\frac{\varDelta_{,r}}{2\varDelta}-\frac{1}{r}\Big) \Big]\notag\\
&+\frac{\varDelta}{r^2}\Big[-\Big(\frac{2}{r^2}-\frac{4}{r^2}\Big)+\frac{1}{2}\Big(-\frac{2}{r^2}-\frac{\varDelta_{,rr}}{\varDelta}+\frac{(\varDelta_{,r})^2}{\varDelta^2}\Big)\notag\\
&+\frac{2}{r}\Big(\frac{1}{r}-\frac{\varDelta_{,r}}{2\varDelta}\Big)-(\frac{\varDelta_{,r}}{2\varDelta}-\frac{1}{r})^2-\Big(\frac{1}{r}-\frac{\varDelta_{,r}}{2\varDelta}\Big)^2\notag\\
&-\frac{1}{r^2}-\frac{1}{r^2} \Big]\notag\\
&+\frac{1}{r^2}\Big[\frac{1}{\sin^2\theta}-\frac{\varDelta_{,r}}{r}+\frac{\varDelta}{r^2}+\frac{2}{r}\Big(-\frac{\varDelta}{r}\Big)\notag\\
&-2\Big(-\frac{\varDelta}{r}\Big)\frac{1}{r}-\Big(\frac{\cos\theta}{\sin\theta}\Big)^2 \Big]\notag\\
&+\frac{1}{r^2\sin^2\theta}\Big[\frac{\varDelta}{r^2}\sin^2\theta-\cos^2\theta+\sin^2\theta+\frac{2}{r}\Big(-\frac{\varDelta}{r}\Big)\notag\\
&+\frac{\cos\theta}{\sin\theta}(-\sin\theta \cos \theta)-\Big(-\frac{\varDelta}{r}\Big)\frac{1}{r}-(-\sin\theta \cos \theta)\frac{\cos\theta}{\sin\theta}\notag\\
&-\frac{1}{r}\Big(-\frac{\varDelta}{r}\Big)-\frac{\cos\theta}{\sin\theta}(-\sin\theta \cos \theta) \Big]\notag \\
&=\frac{2-\varDelta_{,rr}}{r^2} .
\end{align}
Taking a contraction of Einstein's equations, we get
\begin{align}
R^{\mu}_{\mu}-\frac{1}{2}\delta^{\mu}_{\mu}R=\kappa \notag T^{\mu}_{\mu}   \notag  \\
R-2R=\kappa T \notag \\
-R=\kappa T .
\end{align}
Using this, in scalar tensor theory, we have the effective mass resulting from the coupling of matter and a scalar field is,
\begin{align}
\mu^2_s&=\mu_s^2(r,\theta) \notag\\
            &=-2\alpha T(r,\theta) \notag \\
            &=2\alpha R(r,\theta) \notag\\
            &=2\alpha \frac{2-\varDelta_{,rr}}{r^2}+O(a^2) \\
            &\rightarrow 2\alpha \frac{2-\varDelta_{,rr}}{r^2} .
\end{align}

\section{\label{derivationKG}Klein-Gordon equation}
The Klein-gordon equation in a rotating BH spacetime surrounded by a DM halo is
\begin{multline}
-\frac{A}{\varDelta \Sigma^2}\partial^2_t \Psi+\frac{ar(-2M-r+rf(r))}{\varDelta \Sigma^2}\partial_t \partial_{\varphi}\\
+\frac{1}{\Sigma^2}\partial_{\theta}(\sin\theta \partial_{\theta})\Psi+\frac{1}{\Sigma^2}\partial_r(\varDelta \partial_r)\Psi \notag \\
+\frac{ar(-2M-r+rf(r))}{\varDelta \Sigma^2}\partial_{\varphi} \partial_t \\
+\frac{-2Mr-r^2+\Sigma^2+r^2f(r)}{\varDelta \Sigma^2 sin^2 \theta}\partial_{\varphi}^2 \Psi \\
=\mu_s^2(r,\theta)\Psi .
\end{multline}
Adopting the ansatz of separation of variables,
\begin{equation}
\Psi=e^{-i \omega t}e^{-i m \varphi}R(r)S(\theta) ,
\end{equation}
we get 
\begin{multline}
\frac{A\omega^2}{\varDelta \Sigma^2}\Psi+\frac{am\omega(\varDelta -a^2-r^2)}{\varDelta \Sigma^2}\\
+\frac{1}{\Sigma^2 \sin\theta}\partial_{\theta}(\sin \theta \partial_{\theta})\Psi+\frac{1}{\Sigma^2}\partial_r(\varDelta \partial_r)\Psi \\
+\frac{am\omega(\varDelta -a^2-r^2)}{\varDelta \Sigma^2}+\frac{\varDelta-a^2 \sin^2\theta}{\varDelta \Sigma^2\sin^2\theta}(-m^2)\Psi \\
=\mu_s^2(r,\theta)\Psi .
\end{multline}
By sides multipling $\Sigma^2$,and taking the limit of $a^2\rightarrow0$, we have $\Sigma^2\mu_s^2(r,\theta)\rightarrow r^2\mu^2_s(r)$ and,
\begin{align}
&\Big[ \frac{(r^2+a^2)}{\varDelta}-a^2 \sin\theta \Big]\omega^2\Psi+\frac{2am\omega(\varDelta -a^2-r^2)}{\varDelta }\notag\\
&+\frac{1}{ \sin\theta}\partial_{\theta}(\sin \theta \partial_{\theta})\Psi+\partial_r(\varDelta \partial_r)\Psi 
-\frac{1}{\sin^2\theta}\psi+\frac{a^2m^2}{\varDelta}\psi \notag\\
&=r^2\mu_s^2(r) ,
\end{align}
\begin{align}
&\partial_r(\varDelta \partial_r)\Psi+\Big[\frac{\omega^2(r^2+a^2)^2-2am\omega(r^2+a^2)+a^2\omega^2}{\varDelta}\notag\\
&-(a^2\omega^2-2am\omega+r^2\mu_s^2(r))\Big]\Psi \notag\\
&=-\frac{1}{ \sin\theta}\partial_{\theta}(\sin \theta \partial_{\theta})\Psi-a^2\omega^2\cos^2\theta\Psi+\frac{m^2}{\sin^2\theta}\Psi .
\end{align}
Since both sides are equal, these should be constant, which we denote as $\Lambda_{\ell m}$,\par
Finally we get angular equation
\begin{equation}
\frac{1}{\sin\theta}\partial_{\theta}(\sin\theta \partial_{\theta})S(\theta)+(\Lambda_{\ell m}+a^2\omega^2\cos^2\theta-\frac{m^2}{\sin^2\theta})S(\theta)=0 ,
\end{equation}
and radial equation
\begin{align}
\varDelta \partial_r (\varDelta \partial_r)R(r)+[\omega^2(r^2+a^2)^2-2am\omega(r^2+a^2)+a^2 m^2\notag\\
-(\Lambda_{\ell m}+a^2 m^2-2am\omega)\varDelta+\alpha\varDelta^2 \varDelta_{,rr}]R(r)=0 .
\end{align}

\bibliography{DMHST.bib}

\end{document}